\documentclass[aps,prd,floatfix,showpacs,showkeys,superscriptadress,unsortedaddress,nofootinbib,onecolomn]{revtex4-2}
\usepackage{amsmath,mathtools,amssymb,bm,bbm,slashed}
\usepackage[left=2cm,right=2cm]{geometry}
\usepackage{hyperref}
\hypersetup{colorlinks=true, linkcolor=blue, citecolor=blue, urlcolor=blue}
\usepackage{graphicx, float}
\usepackage{enumerate,paralist}
\usepackage{braket}
\usepackage{color}
\usepackage{todonotes}
\usepackage{upgreek}
\usepackage{tikz}
\usepackage{caption}
\usepackage{subcaption}
\usepackage{multirow}
\usepackage{tabularray}

\newcommand{\nn}{\nonumber \\}
\newcommand{\sF}{\scriptscriptstyle{f}}
\newcommand{\sff}{\scriptscriptstyle{(f)}}

\newcommand{\shp}{\shortparallel}
\newcommand{\sperp}{\scriptscriptstyle{\perp}}
\newcommand{\splus}{\scriptscriptstyle{(+)}}
\newcommand{\sminus}{\scriptscriptstyle{(-)}}
\begin{document}
%\color{white}
%\pagecolor{black!100}
\title{\textbf{Quark propagator and di-lepton production rate in a hot, dense and very strongly magnetized rotating Quark-Gluon Plasma}}
\author{Aritra Das}
\email[Email: ]{arisbho2007das@gmail.com}
\thanks{Official email: \href{aritra.das@niser.ac.in}{aritra.das@niser.ac.in}}
\affiliation{School of Physical Sciences, National Institute of Science Education and Research,
		An OCC of Homi Bhabha National Institute, Jatni-752050, India}
\begin{abstract}
		In this paper, a theoretical calculation of thermal di-lepton production rate is reported from a hot and dense, unbounded rotating quark gluon plasma in the presence of very strong uniform background magnetic field typically generated at heavy-ion collision experiments. In this extreme magnetic field, the quarks as well as anti-quarks are approximated to be confined in the lowest Landau level (LLL). Firstly, I have converted the expression of the quark propagator in LLL approximation to the momentum space representation. After that, using the derived quark propagator, the di-lepton production rate (DPR) is calculated from the photon polarization tensor (PPT) with the help of the framework of thermal field theory. The system is first confined in a cylinder of radius $R_{\sperp}$ and then $R_{\sperp}$ is taken to be very large in order to consider the situation of unbounded system. Depending on the charge of participating quarks, the role of rotation is to alter the chemical potential. The DPR is suppressed prominently for low invariant mass with respect to the LLL-approximated non-rotating case that was reported earlier.        
	\end{abstract}
\maketitle
%----------------------------------------
%.      section
%----------------------------------------
\section{Introduction}
% HIC intro
In the peripheral relativistic heavy ion collisions, a tremendous amount of magnetic field is generated in the direction perpendicular to the reaction plane due to the relative motions of the participating ions. Theoretical calculations as well as numerical simulations indicate that the generated magnetic field, whose strength reaches upto $B \sim 10^{18}$G at RHIC and $B\sim 10^{19}$G at LHC~\cite{McLerran:2013hla,Skokov:2009qp,Tuchin:2012mf}, is one of the highest among all produced in the Universe. 
% B phenomena
Recent studies on the effect of magnetic field on color-deconfined quark-gluon plasma indicates the emergence of several novel phenomena like chiral magnetic effect~\cite{Fukushima:2008xe,Kharzeev:2013ffa}, chiral vortical effect~\cite{Kharzeev:2015znc}, chiral magnetic wave~\cite{Kharzeev:2010gd}, magnetic~\cite{Gusynin:1994re,Gusynin:1995nb,Shovkovy:2012zn} and inverse-magnetic catalysis~\cite{Bali:2012zg,Bruckmann:2013oba,Andersen:2014xxa}, superconductivity of QCD vacuum~\cite{Chernodub:2010qx,Chernodub:2011mc} and many more.\\

% \Omega discovery and possible implications of it
Apart from the strong magnetic field, a high vorticity is also present in the created medium. It happens due to the conversion of large angular momentum, carried by the initial colliding nuclei, to vorticity as a result of the conservation of angular momentum.
The generated vorticity, calculated to be $\sim 10^{22}\;\text{s}^{-1}$~\cite{Becattini:2015ska}, induces a spin polarization~\cite{Becattini:2007sr, Becattini:2013fla, Liang:2004xn}, directed perpendicular to the reaction plane, when averaged over the whole fluid. Recent experimental observation of $\Lambda$ and $\overline{\Lambda}$-hyperon spin polarization~\cite{STAR:2017ckg} by STAR collaboration has given a new direction to this already matured field of research. This experimental observation along with some novel theoretical findings, like rotational suppression of chiral condensate~\cite{Jiang:2016wvv}, further boosted interest in the heavy-ion community to re-investigate various phenomena. For example, in Ref.~\cite{Chernodub:2016kxh,Chernodub:2017ref,Wang:2018sur,Zhang:2020hha} chiral phase transition is investigated in the ambit of effective model under rotation. Next, the investigation of QCD phase transitions under rotation gained the attention of the lattice QCD community after the formulation of lattice Quantum Chromodynamics (LQCD) in a rotating frame reported in~\cite{Yamamoto:2013zwa}. Thereafter, using LQCD simulations, the confinement deconfinement transitions~\cite{Braguta:2020biu,Braguta:2021jgn,Yang:2023vsw}, equation of state~\cite{Braguta:2023kwl} in rotating QCD medium was studied. Moreover, shear~\cite{Satapathy:2023cru} and bulk~\cite{Satapathy:2023oym} viscosities were studied in a while ago.\\

The combined effect of magnetic field with rotation showed ``rotational magnetic inhibition"~\cite{Chen:2015hfc} where by increasing $\Omega$, the amount of suppression of the dynamical mass with increasing the background B-field was enhanced. This conclusion was confirmed for non-homogeneous condensate by taking into account 1-flavor~\cite{Sadooghi:2021upd} NJL model and 2-flavor~\cite{Mehr:2022tfq} NJL model.\\

% need to consider em probes
Nowadays, there is growing interest in probing the properties of QGP, after witnessing the emergence of novel phenomena in the these extreme environments. Electromagnetic probes, like di-leptons, real photons, are considered one of the cleanest among all the probes of QGP.  Once produced, they interact very feebly $(\alpha_s \gg \alpha_{\textsf{EM}})$ with the strongly interacting plasma due to their electromagnetic nature.    
%past works in this direction
The investigation of thermal dilepton production rate (DPR) in the presence of background magnetic field was reported in Ref.~\cite{Sadooghi:2016jyf} using Ritus eigenfunction method. Also, several calculations were carried out by restricting the strength of the field, e.g. in strong field~\cite{Bandyopadhyay:2016fyd} (the case of lowest Landau level), weak field~\cite{Bandyopadhyay:2017raf}. In Ref.~\cite{Ghosh:2018xhh}, authors calculated DPR by considering arbitrary B-field but restricted the component of di-lepton momentum along the direction of B-field, i.e. for $p_{\sperp}=0$ but $p_z\neq  0$, only. This restriction was later lifted in Ref.~\cite{Das:2021fma,Wang:2022jxx} by including $p_{\sperp}\neq 0$ in the calculation.\footnote{Authors of Ref.~\cite{Hattori:2020htm}, also calculated DPR in magnetised medium taking into account all LL and argued that the energy momentum distribution of di-leptons can indicate to vacuum dichroism and vacuum birefringes.} Also the production rate of real photons~\cite{Wang:2020dsr,Wang:2021ebh} along with (photon + dilepton) emission anisotropy~\cite{Wang:2023fst} was also studied in recent times. In a non-magnetized medium and in the presence of a background rotation, prompt photon spectrum~\cite{Buzzegoli:2023scn} and thermal dilepton rate~\citep{Wei:2021dib} was reported of late. As for the production rate of electromagnetic probes in hot, dense, magnetorotating QGP medium is concerned, there is a beautiful work~\cite{Buzzegoli:2023vne} that investigated the prompt photon spectrum.	\\
 
% what are there in various sections
In this work, I have calculated the DPR from a rotato-magnetic QGP medium in LLL approximation. In Sec.~\ref{sec:setup}, the technique of incorporating the effect of rotation in the background field is introduced. In Sec.~\ref{sec:quark_prop}, I derive the expression of quark propagator in a strong magnetic field under rotation. Next, in Sec.~\ref{sec:PPT}, the imaginary part of Lorentz-traced photon polarization tensor is computed under LLL approximation. The exact analytic expression of DPR is obtained in Sec.~\ref{sec:DPR}. In Sec.~\ref{sec:discussions}, I have tried to interpret the obtained expression of the rate along with some approximations and limitations involved. Next, I have studied the behavior of the rate by varying various parameters of the expression in Sec.~\ref{sec:results}. In Sec.~\ref{sec:conclusion}, I conclude and put forward some suggestions to further improve the work.   
%----------------------------------------
%.      section
%----------------------------------------
\section{Setup}\label{sec:setup}
In this section, I discuss about Dirac equation for a quark in the presence of an uniform space-time independent background magnetic field with magnitude $B$ in a reference frame rotating with an angular velocity $\Omega$. Keeping in mind the situation created in HIC experiments, both the magnetic field $B$ and the angular velocity $\Omega$ are taken along the same direction. Along that direction, I orient the $z$-axis of the coordinate system. Also, the direction of propagation of the two participating ions is taken along the $x$-axis. In this case, the metric appropriate to the rotating frame is written as
\begin{align}
g_{\mu\nu} = \begin{pmatrix}
1-(x^2+y^2)\Omega^2 & y\Omega & -x\Omega & 0 \\
y\Omega & -1 & 0 & 0 \\
-x\Omega & 0 & -1 & 0 \\
0 & 0 & 0 & -1 
\end{pmatrix}. \label{eq:rotating_metric}
\end{align}  
The Dirac equation, describing a quark of flavor $f$ in a rotating frame, is written as 
\begin{align}
\left[i \gamma^{\mu} \left(\partial_{\mu} + \Gamma_{\mu}\right) - m_f\right]\psi = 0, \label{eq:Dirac_rotating}
\end{align}
where $m_f$ is the mass of the quark and $\Gamma^{\mu}$ is the affine connection which is defined by 
\begin{align}
\Gamma^{\mu} = - \frac{i}{4}\omega_{\mu\hat{i}\hat{j}}\sigma^{\hat{i}\hat{j}}, \label{eq:connection_def}
\end{align}
where Latin indices with hat, i.e., $\hat{i},\hat{j}, \ldots = \hat{t}, \hat{x}, \hat{y}, \hat{z}$, denote the directions of the Cartesian coordinate system in the laboratory frame and Greek indices $\mu, \nu, \ldots = t, x, y, z$ denote the general coordinates in the rotating frame. The quantities $\omega_{\mu\hat{i}\hat{j}}$, $\sigma^{\hat{i}\hat{j}}$ in Eq.~\eqref{eq:connection_def}, are defined by
\begin{align}
\omega_{\mu\hat{i}\hat{j}} \equiv g_{\lambda\rho}e^{\lambda}_{\hat{i}}\left(\partial_{\mu}e^{\rho}_{\hat{j}} + \Gamma^{\rho}_{\mu\nu}e^{\nu}_{\hat{j}}\right), \qquad \sigma^{\hat{i}\hat{j}} = \frac{i}{2}\left[\gamma^{\hat{i}}, \gamma^{\hat{j}}\right], \label{eq:omega_sigma}
\end{align}
where  
\begin{align}
\Gamma^{\lambda}_{\mu\nu} = \frac{1}{2}g^{\lambda\sigma}\left(\partial_{\mu}g_{\sigma\nu}+\partial_{\nu}g_{\mu\sigma}-\partial_{\sigma}g_{\mu\nu}\right) \label{def:christoffel_metric}
\end{align}
is the Christoffel connection and $e^{\mu}_{\hat{i}}$ is the vierbein that connects the general coordinates with the Cartesian coordinate is written as $x^{\mu} = e^{\mu}_{\hat{i}} x^{\hat{i}}$. In this case, the components of vierbein are given by 
\begin{align}
e^{t}_{\hat{t}} = e^{x}_{\hat{x}} = e^{y}_{\hat{y}} = e^{z}_{\hat{z}} = 1, \quad e^{x}_{\hat{t}} = y\Omega, \quad e^{y}_{\hat{t}} = -x\Omega, \label{eq:unit_vec_comp}
\end{align} with other components being zero. For the Christoffel connection, the non-zero components are written as 
\begin{align}
\Gamma^{x}_{ty} = \Gamma^{x}_{yt} = - \Omega, \quad \Gamma^{y}_{tx} = \Gamma^{y}_{xt} = \Omega, \quad \Gamma^{x}_{tt} = - x\Omega^2, \quad \Gamma^{y}_{tt} = - y\Omega^2. \label{eq:Christoffel_comp}
\end{align} 
With this, the only component of affine connection that survives is given as
\begin{align}
\Gamma_{t} = -\frac{i}{2}\Omega\sigma^{\hat{x}\hat{y}}. \label{eq:only_surving_christoffel}
\end{align}
Now the components of the gamma matrices in the rotating frame are given by $\gamma^{\mu} = e^{\mu}_{\hat{i}}\gamma^{\hat{i}}$. In this case, one has
\begin{align}
\gamma^{t} = \gamma^{\hat{t}},\quad \gamma^{x} = y\Omega\gamma^{\hat{i}} + \gamma^{\hat{x}}, \quad \gamma^{y} = -x\Omega\gamma^{\hat{t}} + \gamma^{\hat{y}}, \quad \gamma^{z} = \gamma^{\hat{z}}. \label{eq:gamma_matrix_comp}
\end{align} 
Now, in the presence of background magnetic field, the partial derivatives $\partial_{\mu}$ are replaced by covariant derivatives $D_{\mu}$ as $\partial_{\mu} \rightarrow D_{\mu} = \partial_{\mu} - iq_f \mathcal{A}_{\mu}$, where $\mathcal{A}_{\mu}$ is the gauge field responsible for the generation of the background magnetic field. It can be chosen in various gauges, but for our purpose I choose it in the symmetric gauge where $\mathcal{A}^{\mu} = \dfrac{B}{2}(0,-y, x, 0)$. Also $q_f$ is the charge of quark of flavor $f$. In this paper, I consider only light $u$ and $d$ quarks whose charges are given by $q_u = \dfrac{2}{3}e$ and $q_d=-\dfrac{1}{3}e$, respectively. Here $e$, taken as the absolute value of electron's charge, is positive.  Thus, the general form of the Dirac equation describing rotating fermions in the magnetized QGP is given as
\begin{align}
\left[i\gamma^{\mu}\left(D_{\mu} + \Gamma_{\mu}\right) - m_f \right]\psi = 0.
\end{align}
In the case of uniform rotation $\bm{\Omega} = \Omega \bm{\hat{z}}$,  and magnetic field $\bm{B} = B \bm{\hat{z}}$, it becomes
\begin{align}
\left[i\gamma^{\hat{t}}\left(\partial_t+\Omega y\partial_x - \Omega x\partial_y-\frac{i}{2}\Omega\sigma^{\hat{x}\hat{y}}\right)+i\gamma^{\hat{x}}\left(\partial_x+i\frac{q_fB}{2}y\right) + i\gamma^{\hat{y}}\left(\partial_{y}-i\frac{q_fB}{2}x\right) + i\gamma^{\hat{z}}\partial_z-m_f\right]\psi
 = 0.
\end{align}
Note that I have retained gamma matrices in the laboratory frame but coordinates in the rotating frame for the convenience of our calculation. From now on, I simply write $\gamma^{\hat{t}}, \gamma^{\hat{x}}, \cdots$ as $\gamma^0, \gamma^1, \cdots$. In this way 
\begin{align}
\sigma^{\hat{x}\hat{y}} = \sigma^{12} = \begin{pmatrix}
\sigma_z && 0 \\
0 && \sigma_z
\end{pmatrix},
\end{align}
where $\sigma_z$ is the $z$ component of the Pauli spin matrix.
%----------------------------------------
%.      section
%----------------------------------------
\section{The Quark Propagator}\label{sec:quark_prop}
This section deals with the expression of the quark propagator in the rotato-magnetic background that is necessary for the calculation of DPR. I shall quote the expression of the quark propagator in a magnetised rotating medium in subsec.~\ref{subsec:quark_prop_gen} of this section. Then, in subsec.~\ref{subsec:prop_LLL}, I shall represent it in a suitable form for my calculation of DPR under LLL approximation valid for in the case of very strong B-field.  
%----------------------------------------
%.      subsection
%----------------------------------------
\subsection{The Quark Propagator in a bounded magneto-rotating system}\label{subsec:quark_prop_gen}
The equation that the quark propagator of flavor $f$ satisfies is given as
\begin{align}
&\bigg[i\gamma^{0}\left(\partial_t+\Omega y\partial_x - \Omega x\partial_y-\frac{i}{2}\Omega\sigma^{12}\right)+i\gamma^{1}\left(\partial_x+i\frac{q_fB}{2}y\right) + i\gamma^{2}\left(\partial_{y}-i\frac{q_fB}{2}x\right)  + i\gamma^{3}\partial_z-m_f\Bigg]S^{\sff}_{\alpha\beta}(\mathcal{X},\mathcal{X}^{\prime})
 \nonumber\\
&\hspace{13cm}= \delta^{(4)}(\mathcal{X}-\mathcal{X}^{\prime})\delta_{\alpha\beta}, \label{eq:prop_eq}
\end{align}  
where I have collectively denoted the space-time coordinates by $\mathcal{X}^{\mu} = (t,x,y,z)=(t,\bm{r})$ and ${\mathcal{X}^{\prime}}^{\mu}=(t^{\prime},x^{\prime},y^{\prime},z^{\prime})=(t^{\prime},\bm{r}^{\prime})$. Before proceeding to the solution of Eq.~\eqref{eq:prop_eq}, one can note that it is convenient to confine the system in a cylinder with radius $R_{\sperp}$ and work with the cylindrical polar coordinates where $x = r_{\sperp}\cos\phi$, $y=r_{\sperp}\sin\phi$, with $t,z$ remaining unchanged. Here $r_{\sperp}=\left\vert\bm{r}\cdot\bm{\hat{z}}\right\vert$ is the distance of point $\bm{r}$ from the axis of rotation. The solution of Eq.~\eqref{eq:prop_eq} was investigated thoroughly in Ref.~\cite{Sadooghi:2021upd}. Below, I just quote the expression
\begin{align}
	&S^{\sff}_{\alpha\beta}(\mathcal{X},\mathcal{X}^{\prime}) \nn
	&= i \sum_{n,\ell}\int\frac{dk_0dk_z}{(2\pi)^2}\mathcal{C}_{n,\ell,s_q}^2\,e^{-ik_0(t-t^{\prime})+ik_z(z-z^{\prime})}
	\left[\mathbbm{P}^{\sff}_{\lambda_n,\ell}(\xi,\phi)\right]_{\alpha\rho}\left(\frac{\gamma\cdot\tilde{k}_{\lambda_n,\ell,+}+m_{\sF}}{(k_0+\Omega j)^2-\epsilon^{\sff\,2}_{\lambda_n,k_z}}\right)_{\rho\lambda}\left[\mathbbm{P}^{\sff}_{\lambda_n,\ell}(\xi^{\prime},\phi^{\prime})\right]^{\dagger}_{\lambda\beta}. \label{eq:prop_BC}
\end{align}
As we can see Eq.~\eqref{eq:prop_BC} involves many terms, which are explained one after another. Here $\alpha$, $\beta$, $\lambda$, $\rho$ denotes Lorentz indices. The term $j$ is a positive or negative half-integer written as
\begin{align}
j \equiv \ell + \frac{1}{2}, \label{eq:jdef}
\end{align}
and $\tilde{k}_{\lambda_n,\ell,\kappa}$ is the \emph{Ritus momentum} which is given by
\begin{align}
\tilde{k}_{\lambda_n,\ell,\kappa} = \left(k_0+\kappa\Omega j,0,\kappa s_{\ell}\sqrt{2\lambda_n|q_fB|},k_z\right),
\end{align} 
with $s_{\ell}=\textsf{sgn}(\ell)$, $\kappa = \pm 1$. Here $\kappa = 1$ and $\kappa = -1$ are related to the particle and antiparticle solution of the Dirac equation in rotato-magnetic medium, respectively.
Also, for any four vector $a^{\mu} = (a_0,a_x,a_y,a_z)$, I define
\begin{align}
&\gamma\cdot a = \slashed{a} = \gamma^0a_0 -\gamma^1a^1-\gamma^2a^2-\gamma^3a^3, \quad a^{\mu}_{\shp} = (a_0,0,0,a_z),\quad a^{\mu}_{\sperp}=(0,a_x,a_y,0) \nn
&\gamma\cdot a_{\shp} = \slashed{a}_{\shp} = \gamma^0 a_0 - \gamma^3 a^3, \quad \gamma\cdot a_{\sperp} = \gamma^1 a^1 + \gamma^2 a^2, \nn
&A^2 = a_0^2 - a_x^2 - a_y^2 - a_z^2, \quad a_{\shp}^{2} = a_0^2-a_z^2, \quad a_{\sperp}^2= a_x^2 + a_y^2,\quad a^2 = |\bm{a}|^2 = a_x^2+a_y^2+a_z^2.
\end{align}
The factor $\mathbbm{P}^{\sff}_{\lambda_n,\ell}(\xi,\phi)$ is given as 
\begin{align}
\mathbbm{P}^{\sff}_{\lambda_n,\ell}(\xi) = \mathit{P}_{+}f^{+}_{\lambda_n,\ell,s_q}(\xi,\phi)+ \mathit{P}_{-}f^{-}_{\lambda_n,\ell,s_q}(\xi,\phi),\qquad\text{with}\quad \xi = \frac{|q_fB|r_{\sperp}^2}{2},
\end{align}
where $\mathit{P}_{\pm} = \dfrac{1}{2}\left(\openone\pm i\gamma^1\gamma^2\right)$ and
\begin{align}
&f^{+}_{\lambda_n,\ell,s_q}(\xi,\phi) = \mathcal{C}^{+}_{n,\ell,s_q}e^{i\ell\phi}\,\,\Phi^{+}_{\lambda_n,\ell,s_q}(\xi),\nn
&f^{-}_{\lambda_n,\ell,s_q}(\xi,\phi) = \mathcal{C}^{-}_{n,\ell,s_q}e^{i(\ell+1)\phi}\,\,\Phi^{-}_{\lambda_n,\ell,s_q}(\xi). \label{eq:fpm_def}
\end{align}
The subscripts and superscripts appearing in $f$ bear different meanings which are explained one after another:
\begin{inparaenum}[(i)]
\item The indices $+$ and $-$ in the superscript of the function $f$ are related to the spin up and spin down of a quark, respectively.
\item $s_q=\textsf{sgn}(q_fB)$, with $\textsf{sgn}$ denoting the sign function. It takes the value of $\pm 1$. In this paper, I have taken the background magnetic field to point along the $z$-direction such that $s_q=+1$ for positively charged particles and $s_q=-1$ for negatively charged particles.
\item $\lambda_n$, with ($n=0,1,2,\cdots$), is a parameter analogous to Landau levels in the case of non-rotating magnetic medium. The method to determine it's value, for system with finite $R_{\sperp}$, is discussed later in this subsection. In the limit of $R_{\sperp}\rightarrow \infty$, it takes the value of positive integer including zero.
\item $\ell$ is the orbital angular momentum quantum number whose range depends on $n$ as shown in table~\ref{tab:l_with_HLL}.
\end{inparaenum} 
The pre-factors $\mathcal{C}^{\pm}_{n,\ell,s_q}$ are defined by
\begin{align}
\mathcal{C}^{\pm}_{n,\ell,s_q} = \left(\frac{|q_fB|}{2\pi\int\limits^{\alpha_b}_{0}d\xi \left[\Phi^{\pm}_{\lambda_n,\ell,s_q}(\xi)\right]^2	}\right)^{1/2}, \label{eq:mathcalCpm_def}
\end{align}
where $\alpha_b \equiv \dfrac{|q_fB|}{2}R_{\sperp}^2$.
The functions $\Phi^{\pm}_{\lambda_n,\ell,s_q}$ appearing in Eq.~\eqref{eq:fpm_def} and in Eq.~\eqref{eq:mathcalCpm_def} are defined by
\begin{align}
&\Phi^{+}_{\lambda_n,\ell,s_q}(\xi) = \frac{1}{|\ell|!}\left(\frac{|q_fB|}{2\pi}\frac{\left(\mathcal{N}^{+}_{\lambda_n,s_q}+|\ell|\right)!}{\mathcal{N}^{+}_{\lambda_n,s_q}!}\right)^{1/2}e^{-\xi/2}\xi^{\frac{|\ell|}{2}}\,\,{_1F_1\left(-\mathcal{N}^{+}_{\lambda_n,s_q};|\ell|+1;\xi\right)},\nn
&\Phi^{-}_{\lambda_n,\ell,s_q}(\xi) = \frac{1}{|\ell+1|!}\left(\frac{|q_fB|}{2\pi}\frac{\left(\mathcal{N}^{-}_{\lambda_n,s_q}+|\ell+1|\right)!}{\mathcal{N}^{-}_{\lambda_n,s_q}!}\right)^{1/2}e^{-\xi/2}\xi^{\frac{|\ell+1|}{2}}\,\,{_1F_1\left(-\mathcal{N}^{-}_{\lambda_n,s_q};|\ell+1|+1;\xi\right)}. \label{eq:mathcal_Phi_pm_def}
\end{align}
The function $_1F_1$ in Eq.~\eqref{eq:mathcal_Phi_pm_def} is the confluent hypergeometric function. The terms $\mathcal{N}^{\pm}_{\lambda_n,s_q}$ are defined as
\begin{align}
\mathcal{N}^{+}_{\lambda_n,s_q} &= \lambda_n + \frac{s_q (\ell + 1) - |\ell| -1}{2},\nn
\mathcal{N}^{-}_{\lambda_n,s_q} &= \lambda_n + \frac{s_q \ell - |\ell + 1| -1}{2}.
\end{align}
As mentioned earlier, the $\lambda_n$ (for $n=0,1,2,\cdots$) are Landau level like numbers ($\lambda_n\in \mathbbm{R}$) when $R_{\sperp}$ is finite. It's values depend on $\ell$ and it is computed from the root of the following equations
\begin{align}
	_1F_1\left(-\lambda-\frac{s_q(\ell+1)-|\ell|-1}{2},|\ell|+1,\alpha_b\right) = 0, \text{  for  } \ell \geq 0, \nn
	_1F_1\left(-\lambda-\frac{s_q \ell-|\ell+1|-1}{2},|\ell+1|+1,\alpha_b\right) = 0,  \text{  for  } \ell \leq -1. \label{eq:lambda_root}
\end{align}
 For $\lambda_n$ that satisfy Eq.~\eqref{eq:lambda_root}, one can show numerically that $\mathcal{C}^{+}_{n,\ell,s_q} = \mathcal{C}^{-}_{n,\ell,s_q} \equiv \mathcal{C}_{n,\ell,s_q}$.
%----------------------------------------
%.      subsection
%----------------------------------------
\subsection{Quark Propagator under the Lowest Landau Level Approximation} \label{subsec:prop_LLL}
In this section, I concentrate on the lowest Landau level (LLL) term of the quark propagator written in Eq.~\eqref{eq:prop_eq} for the case of no boundary ($R_{\sperp}\rightarrow \infty $). Now, one can immediately point out some of the restrictions imposed on parameters $\lambda_n$, $\ell$. 
\begin{enumerate}[(a)]
\item \label{item:type1case} First, for the fermion with no boundary $R_{\sperp}\rightarrow \infty$, $\lambda_n$ must be positive integer including $0$. Otherwise, $f^{\sigma}_{n,\ell,s_q}(\xi\rightarrow\infty,\phi)$ can not be kept finite~\cite{Chen:2015hfc}.
\begin{align}
\lambda_{n} \xrightarrow[]{R_{\sperp}\rightarrow\infty} n, \quad \text{where} \quad n\in (0,\mathbbm{Z}^{+}).
\end{align}
\item \label{item:type2case} The confluent Hypergeometric function $_1F_1(a,b;z)$ yields finite number of terms only when 
\begin{align}
( a < 0 ) \quad \textsf{and} \quad ( b < a \quad \textsf{or} \quad b > 0 ). 
\end{align}
\end{enumerate}
Using conditions given in (\ref*{item:type1case}) and (\ref*{item:type2case}), the allowed values of $\ell$ are determined for each values of $n$ and are as shown in Tab~\ref{tab:l_with_HLL}.
\begin{table}[!h]
\begin{center}
\begin{tabular}{|c|c|c|}
\hline
\multicolumn{2}{|c|}{HLL ($n\geq 1$)} \\
\hline
$s_q=1$ & $\ell = -n, -n+1,\cdots, 0, 1, \cdots$ \\
\hline
$s_q=-1$ & $\ell = n-1, n-2, \cdots, 0, -1, \cdots$  \\
\hline
\end{tabular}
\hspace{2cm}
\begin{tabular}{|c|c|c|}
\hline
\multicolumn{3}{|c|}{LLL ($n = 0$)} \\
\hline
{} & $\sigma = 1$ & $\sigma = -1$ \\
\hline
$s_q=1$ & $\ell = 0, 1, \cdots$ & States not allowed \\
\hline
$s_q=-1$ & States not allowed & $\ell = -1, -2, \cdots$ \\
\hline
\end{tabular}
\end{center}
\caption{Table depicting the dependencies of $\ell$ on spin, $s_q$ for states in LLL and higher Landau levels (HLL). In HLL both spins can propagate in contrast to LLL.}
\label{tab:l_with_HLL}
\end{table} 
\newline
Before proceeding to the discussion of LLL scenario, let us note that after determining the constant $\mathcal{C}^{\pm}_{n,\ell,s_q}$ defined in Eq.~\eqref{eq:mathcalCpm_def}, the function $f^{\pm}_{n,\ell,s_q}$ can be written down for the unbounded case ($R_{\sperp}\rightarrow \infty$) in a compact manner. For this purpose, I introduce
\begin{align} 
\ell_{+}\equiv \ell\quad\text{and}\quad\ell_{-}\equiv \ell + 1,
\end{align}
As a result, the functional form of $f^{\pm}_{n,\ell,s_q}$ can be written as
\begin{align}
f^{\sigma}_{n,\ell,s_q}(\xi,\phi) = \left(\frac{|q_fB|}{2\pi}\frac{\mathcal{N}^{\sigma}_{n,s_q}!}{\left(\mathcal{N}^{\sigma}_{n,s_q}+\left\vert\ell_{\sigma}\right\vert\right)!}\right)^{1/2} e^{i\ell_{\sigma}\phi}\; _1F_1\left(-\mathcal{N}^{\sigma}_{n,s_q},|\ell_{\sigma}|+1,\xi\right),\quad\text{with}\quad \sigma=\pm 1, \label{eq:f_unbounded}
\end{align}
where
\begin{align}
\mathcal{N}^{\sigma}_{n,s_q} = n + \frac{s_q\ell_{-\sigma} - |\ell_{\sigma}| -1}{2},  \qquad \ell_{\sigma} = \ell + \delta_{\sigma, -1}.
\end{align}
The $\ell_{\pm}$ are related to $j$ via the relation $j=\ell_{+}+\dfrac{1}{2}=\ell_{-}-\dfrac{1}{2}$.\\

I again emphasize that (i) $\sigma = \pm 1$ is the spin of fermion, (ii) $n$ is the landau level index, (iii) $\ell$ is the orbital angular momentum and (iv) $s_q = \textsf{sgn}(q_fB)$.
For LLL, the allowed values $\ell$ are given by
\begin{align}
\ell = \begin{cases}
0, 1, 2, \cdots, &\quad \text{when} \quad s_q = +1, \\
-1, -2, -3, \cdots, &\quad \text{when} \quad s_q = -1,
\end{cases} \quad\quad \text{for}\quad n=0.
\end{align}
Moreover, using the fact that only $f^{+}_{0,\ell,+1}$ and $f^{-}_{0,\ell,-1}$ survives among all other terms and $[\mathit{P}_{\pm},\gamma_{\shp}] = 0$, one can write the quark propagator in the LLL, from Eq.~\eqref{eq:prop_eq} and Eq.~\eqref{eq:f_unbounded}, as
\begin{align}
&S^{\textsf{LLL}}_{f,\Omega, s_q}(t-t^{\prime},r_{\sperp},\phi-\phi^{\prime},z-z^{\prime}\,;\,0,r^{\prime}_{\sperp},0,0) = i\frac{|q_fB|}{2\pi} e^{-\frac{\xi+\xi^{\prime}}{2}} \int\!\frac{dk_0dk_z}{(2\pi)^2} e^{-i\left[k_{0}(t-t^{\prime})-k_{z}(z-z^{\prime})\right]}\mathcal{I}^{(s_q)}(k_0,k_z;\xi,\xi^{\prime}),\label{eq:LLL_prop1}
\end{align}
where 
\begin{align}
&\mathcal{I}^{(s_q)}(k_0,k_z;\xi,\xi^{\prime})= \displaystyle{\begin{cases}
\sum\limits_{\ell = 0}^{\infty}\frac{\left(\xi\xi^{\prime}\right)^{\frac{\ell}{2}}}{\ell !}e^{i\ell (\phi-\phi^{\prime})}\frac{(k_0+j\Omega)\gamma^0-k_z\gamma^3+m_f}{\left(k_0+j\Omega\right)^2-k_z^2-m_f^2}\mathit{P}_{+}, & \text{for}\quad s_q=1, \\
\sum\limits_{\ell = -1}^{-\infty}\frac{\left(\xi\xi^{\prime}\right)^{-\frac{\ell+1}{2}}}{(-\ell-1) !}e^{i(\ell+1) (\phi-\phi^{\prime})}\frac{(k_0+j\Omega)\gamma^0-k_z\gamma^3+m_f}{\left(k_0+j\Omega\right)^2-k_z^2-m_f^2}\mathit{P}_{-}, & \text{for}\quad s_q=-1,
\end{cases}}\label{eq:Is_part}
\end{align}
where is $j$ defined by Eq.~\eqref{eq:jdef}. 
Next changing the summation index $\ell\rightarrow -\ell -1$ for $s_q=-1$ case in Eq.~\eqref{eq:Is_part}, I can write the quark propagator displayed in Eq.~\eqref{eq:LLL_prop1} in a compact form as
\begin{align}
&S^{\textsf{LLL}}_{f,\Omega, s_q}(t-t^{\prime},r_{\sperp},\phi-\phi^{\prime},z-z^{\prime}\,;\,0,r^{\prime}_{\sperp},0,0) = i\frac{|q_fB|}{2\pi} e^{-\frac{\xi+\xi^{\prime}}{2}} \nonumber \\
&\times \int\!\frac{dk_0dk_z}{(2\pi)^2} e^{-i\left[k_{0}(t-t^{\prime})-k_{z}(z-z^{\prime})\right]}\sum_{\ell = 0}^{\infty}\frac{\left(\xi\xi^{\prime}\right)^{\frac{\ell}{2}}}{\ell !}e^{is_q\ell (\phi-\phi^{\prime})}\frac{\left[k_0+s_q\Omega\left(\ell+\frac{1}{2}\right)\right]\gamma^0-k_z\gamma^3+m_f}{\left[k_0+s_q\Omega\left(\ell+\frac{1}{2}\right)\right]^2-k_z^2-m_f^2}\mathit{P}_{s_q}, \label{eq:prop_LLL_Exact1}
\end{align}
where \begin{align}
\mathit{P_{s_q}} = \frac{1}{2}\left(\openone \pm s_q\,i\gamma^1\gamma^2\right).
\end{align}
Note that the purpose of the variable change $\ell \rightarrow -\ell - 1$ for the summation in case of $s_q=-1$ is to write the sum over $\ell$ to run from $0$ to $\infty$ for both $s_q=\pm 1$. Now, I make a change of variable $\tilde{k}_{0} = k_0 + s_q\left(\ell+\frac{1}{2}\right)\Omega$ after exploiting the translational invariance by defining $\bar{t}=t-t^{\prime}$, $\bar{z} = z - z^{\prime}$ and $\bar{\phi} = \phi - \phi^{\prime}$ coordinate to get
\begin{align}
&S^{\textsf{LLL}}_{f,\Omega,s_q}(\bar{t},r_{\sperp},\bar{\phi},\bar{z};r^{\prime}_{\sperp})= \nonumber\\
&i\frac{|q_fB|}{2\pi}\sum_{\ell = 0}^{\infty}e^{-\frac{\xi+\xi^{\prime}}{2}}\frac{\left(\xi\xi^{\prime}\right)^{\frac{\ell}{2}}}{\ell !}e^{is_q\ell(\bar{\phi} + \Omega \bar{t})}e^{is_q\frac{\Omega}{2}\bar{t}}\int\!\frac{d\tilde{k}_{0}dk_z}{(2\pi)^2}\frac{\tilde{k}_0\gamma^0-k_z\gamma^3+m_f}{\tilde{k}^2_0-k_z^2-m_f^2}e^{-i(\tilde{k}_0\bar{t}-k_z\bar{z})}\mathit{P}_{s_q}.
\end{align}
Now, the sum over $\ell$ can be performed by using the identity
\begin{align}
\sum_{\ell=0}^{\infty}(\xi\xi^{\prime})^{\ell/2}e^{is_q(\phi+\Omega t)\ell}\frac{1}{\ell !} = \exp\left(\sqrt{\xi\xi^{\prime}}e^{is_q(\phi+\Omega t)}\right),
\end{align}
which leads to the simplified exact expression of the propagator in LLL as
\begin{align}
&S^{\textsf{LLL}}_{f,\Omega,s_q}\left(\bar{t},r_{\sperp},\bar{\phi},\bar{z};r^{\prime}_{\sperp}\right)= \nonumber\\
&i\frac{|q_fB|}{2\pi}e^{-\frac{\xi+\xi^{\prime}}{2}}\exp\left(\sqrt{\xi\xi^{\prime}}e^{is_q(\bar{\phi}+\Omega \bar{t})}\right)e^{is_q\frac{\Omega}{2}\bar{t}}\int\!\frac{d\tilde{k}_{0}dk_z}{(2\pi)^2}\frac{\tilde{k}_0\gamma^0-k_z\gamma^3+m_f}{\tilde{k}^2_0-k_z^2-m_f^2}e^{-i(\tilde{k}_0\bar{t}-k_z\bar{z})}\mathit{P}_{s_q}.
\end{align}
At this point, I make the approximation $\bar{\phi} + \Omega \bar{t} = 0$. This is a good approximation in the early stage of heavy ion collision where the fermion is totally dragged by the vortical motion~\cite{Ayala:2021osy}. The reason being the absence of radial flow in such an early time.\\
 
Employing this approximation and using 
\begin{align}
\exp\left(-\frac{\xi+\xi^{\prime}}{2}\right)e^{\sqrt{\xi\xi^{\prime}}} = \exp\left[-\frac{\left(\sqrt{\xi}-\sqrt{\xi^{\prime}}\right)^2}{2}\right]
\end{align}
and finally making a change of variable $\tilde{k}_{0}=k_0 + s_q\frac{\Omega}{2}$, the propagator is obtained as 
\begin{align}
&S^{\textsf{LLL}}_{f,\Omega,s_q}\left(\bar{t},\bar{r}_{\sperp},\bar{\phi},\bar{z}\right)=i\frac{|q_fB|}{2\pi}\exp\left(-\frac{|q_fB|}{4}\bar{r}_{\sperp}^2\right)\int\!\frac{d^2k_{\shp}}{(2\pi)^2}\frac{\left(k_0+s_q\frac{\Omega}{2}\right)\gamma^0-k_z\gamma^3+m_f}{\left(k_0+s_q\frac{\Omega}{2}\right)^2-k_z^2-m_f^2}e^{-i(k_0\bar{t}-k_z\bar{z})}\mathit{P}_{s_q},
\end{align}
where $\bar{r}_{\sperp} = r_{\sperp} - r^{\prime}_{\sperp}$.
Now the Fourier transformation can be performed as the propagator depends on the relative coordinates
\begin{align}
S_{f,\Omega,s_q}^{\textsf{LLL}}(P) = \int\!d^4\bar{\mathcal{X}}\;e^{ip\cdot \bar{\mathcal{X}}}\;S^{\textsf{LLL}}_{f,\Omega,s_q}(\bar{t},\bar{r}_{\sperp},\bar{\phi},\bar{z}),
\end{align}
where $\bar{\mathcal{X}} = \mathcal{X}-\mathcal{X}^{\prime}$.
Writing out the position measure in cylindrical polar coordinates with $\bar{x} = \bar{r}_{\sperp}\cos\bar{\phi}$, $\bar{y}=\bar{r}_{\sperp}\sin\bar{\phi}$ and perpendicular momentum components $p_x = p_{\sperp}\cos\alpha$, $p_y=p_{\perp}\sin\alpha$, one gets 
\begin{align}
S_{f,\Omega,s_q}^{\textsf{LLL}}(P) = i\exp\left(-\frac{p_{\sperp}^2}{|q_fB|}\right)\frac{\left(k_0+s_q\frac{\Omega}{2}\right)\gamma^0-k_z\gamma^3+m_f}{\left(k_0+s_q\frac{\Omega}{2}\right)^2-k_z^2-m_f^2}(\openone + i s_q\gamma^1\gamma^2). \label{eq:prop_mom_LLL}
\end{align}
To arrive at the last expression, I have written $e^{-i\bm{p}_{\perp}.\bar{\bm{r}}_{\sperp}}=e^{-ip_{\perp}\bar{r}_{\sperp}\cos(\bar{\phi}-\alpha)}$ and made use of integration
\begin{align}
&\int\limits_{0}^{2\pi}\!d\bar{\phi}\;\exp\left[ip_{\perp}\bar{r}_{\sperp}\cos(\bar{\phi}-\alpha)\right] = 2\pi J_{0}(p_{\perp}\bar{r}_{\sperp}), \\
&\int\limits_{0}^{\infty}\!d\bar{r}_{\sperp}\;\bar{r}_{\sperp} \exp\left(-\frac{|q_fB|\bar{r}_{\sperp}^2}{4}\right) J_{0}\left(p_{\perp}\bar{r}_{\sperp}\right) = \frac{2}{|q_fB|}\exp\left(-\frac{p_{\perp}^2}{|q_fB|}\right).
\end{align}
Now, noting the appearance of $\Omega/2$ term, one can conclude that the effect of rotation is the same as that of altering the density of the medium by introducing an effective chemical potential $\mu_{\text{eff}} = \pm\Omega/2$ when $s_q=\pm 1$. 
%----------------------------------------
%.      section
%----------------------------------------
\section{Photon Polarization tensor} \label{sec:PPT}
\begin{figure}[!h]
\centering
\includegraphics[scale=0.6]{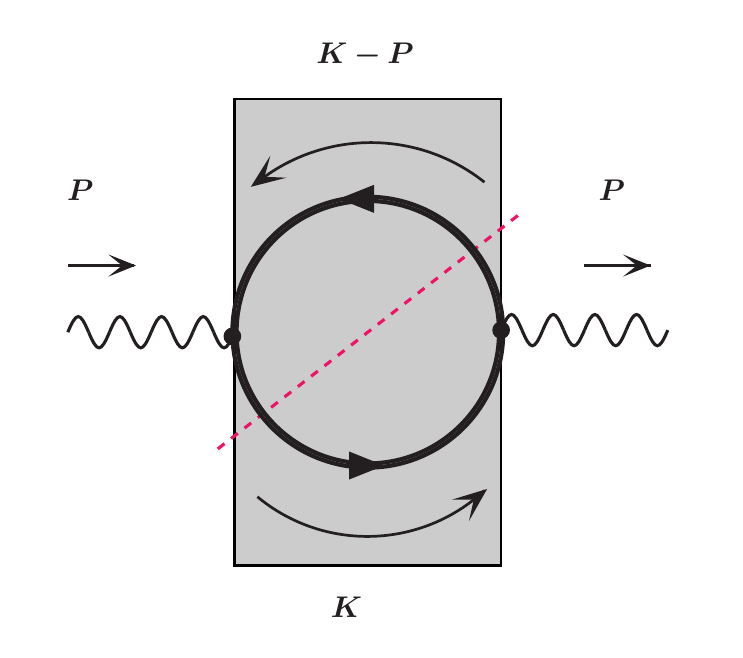}
\caption{Feynman Diagram to compute the one loop photon polarization tensor. The red dashed line indicates the only possible cut allowed to get the quark antiquark annihilation process. The gray background indicates the fact that the quark propagators are affected by the background magnetic field and rotation.}
\label{fig:PPT}
\end{figure}
In this section, I compute the photon polarization tensor (PPT) in the presence of magnetized and rotating QGP. The expression of PPT, reading from the Feynman diagram shown in Fig.~\ref{fig:PPT}, is written as 
\begin{align}
&-i\Pi^{\mu\nu}_{f, B,\Omega,s_q}(p_0,p_z,p_{\sperp}) = N_cq_{f}^2\,iT\!\sum_{k_0}\int\frac{d^3k}{(2\pi)^3}\textsf{Tr}\left[\gamma^{\mu}S^{\textsf{LLL}}_{f,\Omega,s_q}(K)\gamma^{\nu}S^{\textsf{LLL}}_{f,\Omega,s_q}(K-P)\right] \nonumber\\
&=-N_cq_{f}^2\int\frac{d^2\bm{k}_{\bm{\sperp}}}{(2\pi)^2}\exp\left(-\frac{\bm{k}_{\bm{\sperp}}^2+(\bm{k}-\bm{p})_{\mathbbm{\sperp}}^2}{|q_fB|}\right)\,iT\!\sum_{k_0}\int\limits_{-\infty}^{\infty}\frac{dk_z}{2\pi}\textsf{Tr}\left[\gamma^{\mu}\frac{1}{\slashed{\widetilde{k}}_{\shp}+m_f}2\mathit{P}_{s_q}\gamma^{\nu}\frac{1}{\slashed{\widetilde{k}}_{\shp}-\widetilde{p}_{\shp}+m_f}2\mathit{P}_{s_q}\right], \label{eq:PPT_main}
\end{align}
where I have defined
\begin{align}
\widetilde{k}_{\shp}^{\mu} = \left(k_0 + s_q\frac{\Omega}{2}, 0, 0, k_z\right). \label{eq:tilde_def}
\end{align}
The last expression should not be confused with the parallel component of Ritus four-momentum as the latter involves various labels on its subscript. Note that, a replacement 
\begin{align}
\int\limits_{-\infty}^{\infty}\frac{dk_0}{2\pi} \Rightarrow iT\sum_{k_0},\quad\text{where}\quad k_0 = \mu + i(2m+1)\pi T,
\end{align}
for fermions in the loop, according to the prescription of imaginary time formalism (ITF) of thermal field theory, is made~\cite{Mustafa:2022got}. The sum is over discrete Matsubara modes characterized by an integer $m (= 0, \pm 1, \pm 2, \ldots)$. Here $T=\beta^{-1}$ is the temperature $\mu$ is the chemical potential of the plasma. The three-momentum integral measure is split into longitudinal part $d^2\bm{k}_{\sperp}$ and the transverse part $dk_z$. The Dirac trace is performed numerous times in the literature~\cite{Bandyopadhyay:2016fyd,Hattori:2022uzp,Hattori:2022wao} before for non-rotating case. In this calculation, the only difference from the previous one is the replacement $k_0 \rightarrow k_0 + s_q\Omega /2$. The trace is calculated as
\begin{align}
&\mathcal{T}_{\shp}^{\mu\nu} = \textsf{Tr}\left[\gamma^{\mu}\frac{1}{\slashed{\widetilde{k}}_{\shp}+m_f}2\mathit{P}_{s_q}\gamma^{\nu}\frac{1}{\slashed{\widetilde{k}}_{\shp}-\widetilde{p}_{\shp}+m_f}2\mathit{P}_{s_q}\right] \nonumber \\
&=4\left[\widetilde{k}^{\mu}_{\shp}\left(\widetilde{k}^{\nu}_{\shp}-p^{\nu}_{\shp}\right)+\widetilde{k}^{\nu}_{\shp}\left(\widetilde{k}^{\mu}_{\shp}-p^{\mu}_{\shp}\right)-\eta_{\shp}^{\mu\nu}\left\{\left({\eta_{\shp}}\right)_{\lambda\rho}\widetilde{k}^{\lambda}_{\shp}\left(\widetilde{k}^{\rho}_{\shp}-p_{\shp}^{\rho}\right)-m_f^2\right\}\right]. \label{eq:trace}
\end{align}
where $\eta_{\mu\nu} = \left(\eta_{\shp}\right)_{\mu\nu}-\left(\eta_{\sperp}\right)_{\mu\nu} = \textsf{diag}(1,-1,-1,-1)$ is the flat-metric tensor which is broken into parallel and perpendicular part as $\left(\eta_{\shp}\right)_{\mu\nu} = \textsf{diag}(1,0,0,-1)$ and $\left(\eta_{\sperp}\right)_{\mu\nu} = \textsf{diag}(0,1,1,0)$, respectively. \\

The perpendicular integration is performed by using the following integration 
\begin{align}
&\int\limits_{0}^{2\pi}\!d\phi\;\exp\left(\frac{2k_{\sperp}p_{\sperp}}{|q_fB|}\cos\phi\right) = 2\pi I_{0}\left(\frac{2k_{\sperp}p_{\sperp}}{|q_fB|}\right), \label{eq:ang_mom}\\
&\int\limits_{0}^{\infty}\!d k_{\sperp}\; k_{\sperp}I_{0}\left(\frac{2k_{\sperp}p_{\sperp}}{|q_fB|}\right)\exp\left(-\frac{k^2_{\sperp}}{|q_fB|}\right) = \frac{|q_fB|}{4}\exp\left(\frac{p_{\sperp}^2}{2|q_fB|}\right), \label{eq:mag_pmomentum}
\end{align}
and the result is given as 
\begin{align}
\int\frac{d^2\bm{k}_{\bm{\sperp}}}{(2\pi)^2}\exp\left(-\frac{\bm{k}_{\bm{\sperp}}^2+(\bm{k}-\bm{p})_{\mathbbm{\sperp}}^2}{|q_fB|}\right) = \frac{|q_fB|}{8\pi}\exp\left(-\frac{p_{\sperp}^2}{2|q_fB|}\right). \label{eq:perp_int}
\end{align}
The $I_0$ in Eq.~\eqref{eq:ang_mom} and \eqref{eq:mag_pmomentum} is the bessel-$I$ function of the order $0$. It is very evident that the perpendicular momentum is disentangled from parallel momentum in PPT.

Therefore, substituting Eq.~\eqref{eq:perp_int} and Eq.~\eqref{eq:trace} in Eq.~\eqref{eq:PPT_main}, I get
\begin{align}
&-i\Pi^{\mu\nu}_{f, B,\Omega,s_q}(p_0,p_z,p_{\sperp}) = \nonumber \\ &-N_cq_f^2\frac{|q_fB|}{8\pi}e^{-\frac{p_{\sperp}^2}{2|q_fB|}}\,iT\sum_{k_0}\int\limits_{-\infty}^{\infty}\frac{dk_z}{2\pi}\frac{\mathcal{T}^{\mu\nu}_{\shp}}{\left[\left(k_0+s_q\frac{\Omega}{2}\right)^2-k_z^2-m_f^2\right]\left[\left(k_0+s_q\frac{\Omega}{2}-p_0\right)^2-(k_z-p_z)^2-m_f^2\right]}.
\end{align}
%%%%%%%%%%%%%%%%%%%%%%%%%%%%%%%%%%%%%%%%%%

%%% section (dilepton production rate)

%%%%%%%%%%%%%%%%%%%%%%%%%%%%%%%%%%%%%%%%%%%
\section{Dilepton Production Rate} \label{sec:DPR}
Lepton pairs are generated as Landau level states in the rotato-magnetic environment inside the fireball. However, they get converted into the plane wave states $\ket{Q}$, $\ket{Q^{\prime}}$ as soon as they leave the medium. It happens due to the fact that the non-trivial backgrounds under consideration only exist in the formed QGP medium. Now, the probability of re-scattering of lepton pairs inside the plasma is very small. So, one can make an approximation by which all the lepton pairs, produced in the magnetized medium, are converted into plane waves. In this scenario, the matrix amplitude of initial quark and antiquark states $\ket{q(n,\ell,p_z), \overline{q}(n^{\prime},\ell^{\prime},p^{\prime}_z)}$ to get converted into di-leptons is given as $\braket{ Q_1,Q_2 | \mathcal{M} | q(n,\ell,p_z), \overline{q}(n^{\prime},\ell^{\prime},p^{\prime}_z) }$. In this way, the representation of final lepton-antilepton pairs by plane waves is a very good approximation~\cite{Wang:2022jxx}. \\
Also, the lepton antilepton pairs are assumed to be massless $M_l = 0$ because they are the smallest among all the scales involved. Hence, in the scenario in which the quarks move in the strongly magnetized and rotating medium but not the final lepton pairs, the dilepton production per unit space-time and four momentum volume is written as
\begin{align}
	\frac{dN}{d^4\mathcal{X}d^4P} = \frac{\alpha_{\textsf{EM}}}{12\pi^4}\frac{1}{p_0^2-\bm{p}_{\sperp}^2-p_z^2}\frac{\eta_{\mu\nu}}{e^{p_0/T}-1}\sum_{f=u,d}\textsf{Im}\;\Pi^{\mu\nu}_{f, B,\Omega}(p_0,p_z,p_{\sperp}). \label{eq:DR_formula}
\end{align}
Note that, $s_q$ is $+1$ and $-1$ for $q_u$ and $q_d$, respectively. Also, I have $\mathcal{T}_{\shp\;\mu}^{\mu} = \eta_{\mu\nu}\mathcal{T}_{\shp}^{\mu\nu}=8m_f^2$. The frequency sum is performed in Appendix~\ref{app:freq_sum}. Setting $a=\bar{a}=\Omega/2$, $E_1=\sqrt{k_z^2+m_f^2}$, $E_2=\sqrt{(k_z-p_z)^2+m_f^2}$ in Eq.~\eqref{eq:app:freq_I_final} of Appendix~\ref{app:freq_sum}, I get 
\begin{align}
&\eta_{\mu\nu}\Pi^{\mu\nu}_{f, B,\Omega, s_q}(p_0,p_z,p_{\sperp}) = -N_c q_f^2 m_f^2\frac{|q_fB|}{8\pi^2}e^{-\frac{p_{\sperp}^2}{2|q_fB|}}\nonumber\\
&\times\int\limits_{-\infty}^{\infty}d k_z\sum_{r_1=\pm  1}\sum_{r_2=\pm 1}\frac{r_1r_2}{\epsilon_{0,k_z}^{\sff}\epsilon_{0,k_z-p_z}^{\sff}}\frac{1-\widetilde{n}^{(+)}\left(r_1\epsilon_{0,k_z}^{\sff}-s_q\frac{\Omega}{2}\right)-\widetilde{n}^{(-)}\left(r_2\epsilon_{0,k_z-p_z}^{\sff}+s_q\frac{\Omega}{2}\right)}{p_0-r_1\epsilon_{0,k_z}^{\sff}-r_2\epsilon_{0,k_z-p_z}^{\sff} + i\varepsilon}.	\label{eq:Pimumu_sumform}
\end{align}
Introducing $\tilde{n}(x)=1/[\exp (x) + 1]$, Eq.~\eqref{eq:Pimumu_sumform} is re-cast as 
\begin{align}
&\eta_{\mu\nu}\Pi^{\mu\nu}_{f, B,\Omega, s_q}(p_0,p_z,p_{\sperp}) = -N_c q_f^2 m_f^2\frac{|q_fB|}{8\pi^2}e^{-\frac{p_{\sperp}^2}{2|q_fB|}}\nonumber\\
&\times\int\limits_{-\infty}^{\infty}d k_z\sum_{r_1=\pm  1}\sum_{r_2=\pm 1}\frac{r_1r_2}{\epsilon_{0,k_z}^{\sff}\epsilon_{0,k_z-p_z}^{\sff}}\frac{1-\widetilde{n}\left(r_1\epsilon_{0,k_z}^{\sff}-\mu_{s_q}^{\ast}\right)-\widetilde{n}\left(r_2\epsilon_{0,k_z-p_z}^{\sff}+\mu_{s_q}^{\ast}\right)}{p_0-r_1\epsilon_{0,k_z}^{\sff}-r_2\epsilon_{0,k_z-p_z}^{\sff} + i\varepsilon},	\label{eq:Pimumu_sumform_recast}
\end{align}
where an \emph{effective chemical} potential is introduced via
\begin{align}
\mu^{\ast}_{s_q} = \mu + s_q\frac{\Omega}{2}.
\end{align} 
For simplicity, I consider the mass of up and down quark same $m_q\equiv m_u = m_d = 5\;\text{MeV}$. Therefore,
I drop the subscript and/or superscript $f$ in the expression of energy. As I am working in LLL approximation the subscript $0$ denoting LLL is also dropped from now on.\\

The sums over $r_1, r_2$ are performed as
\begin{align}
&\eta_{\mu\nu}\Pi^{\mu\nu}_{f, B,\Omega, s_q}(p_0,p_z,p_{\sperp}) = -N_c q_f^2 m_q^2\frac{|q_fB|}{8\pi^2}e^{-\frac{p_{\sperp}^2}{2|q_fB|}}\nonumber\\
&\times\int\limits_{-\infty}^{\infty} \frac{d k_z}{\epsilon_{k_z}\epsilon_{k_z-p_z}}\Bigg[\frac{1-\widetilde{n}\left(\epsilon_{k_z}-\mu^{\ast}_{s_q}\right)-\widetilde{n}\left(\epsilon_{k_z-p_z}+\mu^{\ast}_{s_q}\right)}{p_0-\epsilon_{k_z}-\epsilon_{k_z-p_z}+i\varepsilon}+\frac{\widetilde{n}\left(\epsilon_{k_z}-\mu^{\ast}_{s_q}\right)-\widetilde{n}\left(\epsilon_{k_z-p_z}-\mu^{\ast}_{s_q}\right)}{p_0-\epsilon_{k_z}+\epsilon_{k_z-p_z}+i\varepsilon}\nonumber\\
&\hspace{2cm}-\frac{\widetilde{n}\left(\epsilon_{k_z}+\mu^{\ast}_{s_q}\right)-\widetilde{n}\left(\epsilon_{k_z-p_z}+\mu^{\ast}_{s_q}\right)}{p_0+\epsilon_{k_z}-\epsilon_{k_z-p_z}+i\varepsilon} -\frac{1-\widetilde{n}\left(\epsilon_{k_z}+\mu^{\ast}_{s_q}\right)-\widetilde{n}\left(\epsilon_{k_z-p_z}-\mu^{\ast}_{s_q}\right)}{p_0+\epsilon_{k_z}+\epsilon_{k_z-p_z}+i\varepsilon}\Bigg].
\end{align}
The imaginary part is calculated by employing the formula
\begin{align}
	\textsf{Im}\frac{1}{u\pm i\varepsilon} = \mp\pi\delta(u).
\end{align}
From here, $B$, $\Omega$ are also dropped from the expression of polarization tensor which is evident from the context. Also, the the presence of $s_q$ is automatically understood if not explicitly specified.
So, one can write 
\begin{align}
&\textsf{Im}\Pi^{\mu}_{f\;\mu} = N_c q_f^2 m_q^2\frac{|q_fB|}{8\pi^2}e^{-\frac{p_{\sperp}^2}{2|q_fB|}}\int\limits_{-\infty}^{\infty} \frac{d k_z}{\epsilon_{k_z}\epsilon_{k_z-p_z}}\Bigg\{\Big[1-\widetilde{n}\left(\epsilon_{k_z}-\mu^{\ast}_{s_q}\right)-\widetilde{n}\left(\epsilon_{k_z-p_z}+\mu^{\ast}_{s_q}\right)\Big]\delta\bigg(p_0-\epsilon_{k_z} \nonumber \\
&-\epsilon_{k_z-p_z}\bigg)+\Big[\widetilde{n}\left(\epsilon_{k_z}-\mu^{\ast}_{s_q}\right)-\widetilde{n}\left(\epsilon_{k_z-p_z}-\mu^{\ast}_{s_q}\right)\Big]\delta\bigg(p_0-\epsilon_{k_z}+\epsilon_{k_z-p_z}\bigg)-\Big[\widetilde{n}\left(\epsilon_{k_z}+\mu^{\ast}_{s_q}\right)-\widetilde{n}\left(\epsilon_{k_z-p_z}\right. \nonumber \\
&\left.+\mu^{\ast}_{s_q}\right)\Big]\delta\bigg(p_0+\epsilon_{k_z}-\epsilon_{k_z-p_z}\bigg)-\Big[1-\widetilde{n}\left(\epsilon_{k_z}+\mu^{\ast}_{s_q}\right)-\widetilde{n}\left(\epsilon_{k_z-p_z}-\mu^{\ast}_{s_q}\right)\Big]\delta\Big(p_0+\epsilon_{k_z}+\epsilon_{k_z-p_z}\Big)\Bigg\}. \label{eq:ImPimumu}
\end{align} 
From the Appendix~\ref{app:kin}, it is evident that there are two kinematic regions that give rise to imaginary part of the potential and leads to various processes. They are 
\begin{enumerate}[(I)]
\item \textbf{Unitary cut: } $\left(\infty > p_0 \geqslant +\sqrt{p_z^2+4m_q^2}\right) \quad \textsf{and}\quad \left(-\infty < p_0 \leqslant -\sqrt{p_z^2+4m_q^2}\right)$,
\item \textbf{Landau cut: } $-|p_z| \leqslant p_0 \leqslant |p_z|$.
\end{enumerate}
\begin{figure}[htbp]
    \centering
       \includegraphics[scale=0.9]{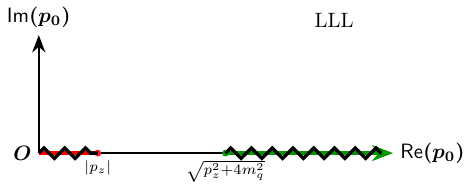}
    \caption{[color online] Kinematic region in $\angle p_0$ plane where $\Pi^{\mu}_{\mu}$ gives non-zero imaginary part. Here only the $\textsf{Re}(p_0) > 0$ portion is shown. The red line denotes Landau cut part which does not contribute to the DPR when both quark and antiquark resides at LLL. However, the green line denotes unitary cut part which do contribute to DPR in the LLL scenario.}
    \label{fig:cut_pics}
\end{figure}
However, the imaginary part arising from Landau cut region does not contribute to any processes when both the quark and/or antiquark reside in LLL.\footnote{Upon lifting the condition of LLL approximation, the Landau cut do contributes to the DPR~\cite{Das:2021fma}} The reason is the unavailability of phase space for producing lepton anti-lepton pairs which is evident from the following argument ---\\
 Let $Q=(q_0, \bm{q})=\left(\sqrt{|\bm{q}|^2+M_l^2},\bm{q}\right)$ and $Q^{\prime} = (q_0^{\prime},\bm{q}^{\prime})=\left(\sqrt{|\bm{q}^{\prime}|^2+M_l^2},\bm{q}^{\prime}\right)$ be the four-momentum of the produced lepton and anti-lepton pair for on-shell final di-leptons, respectively. Then, following the conservation of energy momentum, one has $P = Q + Q^{\prime}$ and $Q^{2} = {Q^{\prime}}^2 = M_l^2$, respectively. Squaring both sides of energy-momentum conservation relation yields 
\begin{align}
 P^2 = Q^2 + {Q^{\prime}}^2 + 2Q\cdot Q^{\prime} = 2 M_l^2 + 2\left(\sqrt{|\bm{q}|^2+M_l^2}\sqrt{|\bm{q}^{\prime}|^2+M_l^2} - \bm{q}\cdot\bm{q}^{\prime}\right).
\end{align}
Lastly, the Schwartz inequality $\sqrt{|\bm{q}|^2+M_l^2}\sqrt{|\bm{q}^{\prime}|^2+M_l^2} - \bm{q}\cdot\bm{q}^{\prime} \geqslant M_l^2$ leads to $P^2=p_0^2-p_{\perp}^2-p_z^2 \geqslant 4M_l^2$, where $M_{l}$ is the mass of leptons (anti-leptons). In the Landau cut region, one has $P^2 = p^2_0-p_z^2-p_{\perp}^2  = -|p_z^2-p_0^2| - p_z^2 \leqslant 0$ leading to the non-availability of phase space as claimed. Moreover, as the condition $p_0 > 0$ and $p_0 < 0$ corresponds to the creation and decay of virtual photons, respectively, I concentrate only on $p_0 > 0$ case. As a result, I only concentrate on $p_0 \geqslant \sqrt{p_z^2+4m_q^2}$ region.\\ 

Among the four terms appearing in the R.H.S of Eq.~\eqref{eq:ImPimumu}, the last one inside $k_z$ integration with $\delta\left(p_0 + \epsilon_{k_z} + \epsilon_{k_z-p_z}\right)$ does not contribute to the production of dileptons. So, I concentrate on the first three terms inside the $k_z$ integral in Eq.~\eqref{eq:ImPimumu}. To perform the $k_z$ integration, I make use of the identity of Dirac delta function 
\begin{align}
\delta(f(u)) = \sum_{i}\left\vert \dfrac{\partial f(u)}{\partial u} \right\vert_{u=u_i}^{-1}\delta(u-u_i),\quad \text{with }u_i\text{ being the }i\text{-th root of }f(u) = 0. \label{eq:delta_id}
\end{align}
The equation $p_0 - \epsilon_{k_z} - \epsilon_{k_z-p_z} = 0$ provides two solution for $k_z$
\begin{align}
k_{z}^{(\pm)} = \frac{1}{2}\left( p_z \pm \frac{\sqrt{p_0^2\,(p_0^2-p_z^2)(p_0^2-p_z^2-4 m_q^2)}}{p_0^2-p_z^2}\right), \label{eq:kz_sol}
\end{align}
for $r_1, r_2 = \pm 1$. The condition $0\leqslant \sqrt{p_z^2+4m_q^2}\leqslant p_0$ further simplifies the expression of $k_z^{(\pm)}$ in Eq.\eqref{eq:kz_sol} to 
\begin{align}
&k^{(\pm)}_{z} = \frac{1}{2}\left( p_z \pm p_0\sqrt{1-\frac{4m_q^2}{p_0^2-p_z^2}}\right).
\end{align}
It leads to 
\begin{align} 
k^{(\pm)}_{z}-p_z = -\frac{1}{2}\left( p_z \mp p_0\sqrt{1-\frac{4m_q^2}{p_0^2-p_z^2}}\right),
\end{align}
with energies
\begin{align}
&\epsilon_{k^{(\pm)}_{z}} = \epsilon^{(\pm)}, \quad 
\epsilon_{k^{(\pm)}_{z}-p_z} =\epsilon^{(\mp)},
\end{align}
where $\epsilon^{(\pm)}$ are defined as
\begin{align}
\epsilon^{(\pm)} \equiv \frac{1}{2}\left(p_0 \pm p_z \sqrt{1-\frac{4m_q^2}{p_0^2-p_z^2}}\right).
\end{align}
%-------------------------------------

%    Subsection 

%--------------------------------------
\subsection{Process - I}
In this case, I concentrate on quark and anti-quark annihilation process $q\overline{q}\rightarrow \gamma^{*}$ where the relevant term is written as 
\begin{align}
&\bigg(\textsf{Im}\Pi^{\mu}_{f\;\mu}\bigg)_{\text{I}} = N_c q_f^2 m_q^2\frac{|q_fB|}{8\pi}\exp\left(-\frac{p_{\sperp}^2}{2|q_fB|}\right) \int\limits_{-\infty}^{\infty}\frac{d k_z}{\epsilon_{k_z}\epsilon_{k_z-p_z}}\Big[1-\widetilde{n}\left(\epsilon_{k_z}-\mu^{\ast}_{s_q}\right)-\widetilde{n}\left(\epsilon_{k_z-p_z}+\mu^{\ast}_{s_q}\right)\Big] \nonumber \\
&\hspace{10cm}\times\delta\bigg(p_0 - \epsilon_{k_z} - \epsilon_{k_z-p_z}\bigg). \label{eq:annImPimumu}
\end{align}
Applying the identity in Eq.~\eqref{eq:delta_id} along with $\epsilon^{(\pm)}>0$\footnote{For the proof of the positivity condition of $\epsilon^{(\pm)}$, see Appendix~\ref{app:kin}} and $p_{\shp}^2 > 0$, the delta function in Eq.~\eqref{eq:annImPimumu} becomes after few steps of algebra,
\begin{align}
\delta\left(p_0-\epsilon_{k_z}-\epsilon_{k_z-p_z}\right) = \frac{2\epsilon^{(+)}\epsilon^{(-)}}{(p_0^2-p_z^2)}\left(1-\frac{4m_q^2}{p_0^2-p_z^2}\right)^{-\frac{1}{2}}\left[\delta \left(k_z - k_z^{(+)}\right) + \delta \left(k_z - k_z^{(-)}\right)\right],
\end{align}
Here, I have used
\begin{align}
\left[ \frac{\partial}{\partial k_z} \delta\left(p_0-\epsilon_{k_z}-\epsilon_{k_z-p_z}\right) \right]_{k_z = k_z^{(\pm)}} = \pm \frac{p_0^2-p_z^2}{2\epsilon^{(\pm)}\epsilon^{(\mp)}}\sqrt{1-\frac{4m_q^2}{p_0^2-p_z^2}}.
\end{align} 
Therefore, an analytic expression of first term is obtained as
\begin{align}
&\bigg(\textsf{Im}\Pi^{\mu}_{f\;\mu}\bigg)_{\text{I}} = N_c q_f^2 m_q^2\frac{|q_fB|}{8\pi}\exp\left(-\frac{p_{\sperp}^2}{2|q_fB|}\right)\frac{2}{(p_0^2-p_z^2)}\left(1-\frac{4m_q^2}{p_0^2-p_z^2}\right)^{-\frac{1}{2}}\Bigg[2-\widetilde{n}\left(\epsilon^{(-)}-\mu_{s_q}^{\ast}\right) \nonumber \\
&\hspace{5cm}-\widetilde{n}\left(\epsilon^{(+)}-\mu_{s_q}^{\ast}\right)-\widetilde{n}\left(\epsilon^{(+)}+\mu_{s_q}^{\ast}\right)-\widetilde{n}\left(\epsilon^{(-)}+\mu_{s_q}^{\ast}\right)\Bigg] \label{eq:t1}
\end{align}
%-------------------------------------

%    Subsection 

%-------------------------------------- 
\subsection{\textbf{Process - II}}
In this case, the corresponding processes are decay processes $q(\overline{q}) \rightarrow q(\overline{q})\gamma^{*}$. The expression relating to this written as
\begin{align}
&\bigg(\textsf{Im}\Pi^{\mu}_{f\;\mu}\bigg)_{\text{II}_{\pm}} = \pm N_c q_f^2 m_q^2\frac{|q_fB|}{8\pi}\exp\left(-\frac{p_{\sperp}^2}{2|q_fB|}\right) \nonumber \\
&\times\int\limits_{-\infty}^{\infty}\frac{d k_z}{\epsilon_{k_z}\epsilon_{k_z-p_z}}\Big\{\left[\widetilde{n}\left(\epsilon_{k_z}\mp\mu^{\ast}_{s_q}\right)-\widetilde{n}\left(\epsilon_{k_z-p_z}\mp\mu^{\ast}_{s_q}\right)\right]\delta\bigg(p_0 \mp \epsilon_{k_z} \pm \epsilon_{k_z-p_z}\bigg)\Bigg\}.
\end{align}
Here the subscript $\text{II}_{+}$ refers to the term with $\delta\left(p_0 - \epsilon_{k_z} +\epsilon_{k_z-p_z}\right)$ and $\text{II}_{-}$ refers to that with $\delta\left(p_0 + \epsilon_{k_z} -\epsilon_{k_z-p_z}\right)$. In this particular kinematic region, I have 
\begin{align}
\left[ \frac{\partial}{\partial k_z} \delta\left(p_0\mp \epsilon_{k_z}\pm\epsilon_{k_z-p_z}\right) \right]_{k_z = k_z^{(\pm)}} = \frac{8m_q^2p_0p_z}{(p_0^2-p_z^2)\epsilon^{(+)}\epsilon^{(-)}},
\end{align}
The delta function is simplified by Eq.~\eqref{eq:delta_id} as
\begin{align}
\delta\left(p_0\mp \epsilon_{k_z}\pm\epsilon_{k_z-p_z}\right) = \frac{\epsilon^{(+)}\epsilon^{(-)}(p_0^2-p_z^2)}{8m_q^2p_0|p_z|}\left[\delta\left(k_z-k^{(+)}_z\right) + \delta\left(k_z-k^{(-)}_z\right)\right].
\end{align}
After performing delta function integration, I arrive at 
\begin{align}
&\bigg(\textsf{Im}\Pi^{\mu}_{f\;\mu}\bigg)_{\text{II}_{\pm}} = \pm N_c q_f^2 m_q^2\frac{|q_fB|}{8\pi}\exp\left(-\frac{p_{\sperp}^2}{2|q_fB|}\right)\frac{\epsilon^{(+)}\epsilon^{(-)}(p_0^2-p_z^2)}{8m_q^2p_0|p_z|} \nonumber \\
&\times\Bigg[\widetilde{n}\left(\epsilon^{(+)}\mp\mu^{\ast}_{s_q}\right)-\widetilde{n}\left(\epsilon^{(-)}\mp\mu^{\ast}_{s_q}\right)+\widetilde{n}\left(\epsilon^{(+)}\mp\mu^{\ast}_{s_q}\right)-\widetilde{n}\left(\epsilon^{(-)}\mp\mu^{\ast}_{s_q}\right)\Bigg] \nonumber \\
&=0 .
\end{align}
Hence, when both of the quark and antiquark resides on LLL, the calculation showed that decay processes are kinematically prohibited in the region where $p_0\geqslant \sqrt{p_z^2+4m_q^2}$. In general, decay processes can never arise from the unitary cut region of the complex $p_0$ plane~\cite{Weldon:1990iw}. Moreover, when both of the quark-antiquark resides on LLL, the Landau cut part does not support the processes where $P^2 < 0$.
%--------------------------------

%  subsection 

%---------------------------------
\subsection{\textbf{The Di-lepton Production Rate}}
The dilepton production per unit four-spacetime volume and per unit four-momentum volume in a hot rotating magnetized medium under LLL approximation is given, upon substituting Eq.~\eqref{eq:t1} in Eq.~\eqref{eq:DR_formula}, as
\begin{align}
&\frac{d\mathcal{N}}{d^4Pd^4\mathcal{X}}\Bigg\vert_{q\bar{q}\rightarrow \gamma^{*}} = N_c\frac{ \alpha^2_{\textsf{EM}} }{12\pi^4}\frac{m_q^2}{p_{\shp}^2 (p_{\shp}^2-p_{\sperp}^2)}{\frac{1}{e^{p_0/T}-1}}\left(1-\frac{4m_q^2}{p_{\shp}^2}\right)^{-\frac{1}{2}}\Bigg\{\left(q_u/e\right)^2|q_uB|\exp\left(-\frac{p_{\sperp}^2}{2|q_uB|}\right)\Bigg[2 \nonumber \\
&\hspace{-0.5cm}-\sum_{I=H,L}\Bigg(\widetilde{n}\left(\epsilon_{I}-\mu^{\ast}_{+}\right)+\widetilde{n}\left(\epsilon_{I}+\mu^{\ast}_{+}\right)\Bigg)\Bigg]+\left(q_d/e\right)^2|q_dB|\exp\left(-\frac{p_{\sperp}^2}{2|q_dB|}\right)\Bigg[2 -\sum_{I=H,L}\Bigg(\widetilde{n}\left(\epsilon_{I}-\mu^{\ast}_{-}\right) \nonumber \\
&\hspace{11cm}+\widetilde{n}\left(\epsilon_{I}+\mu^{\ast}_{-}\right)\Bigg)\Bigg]\Bigg\}. \label{eq:DPR:Final}
\end{align}
In the last line I have introduced two modes, independent of the sign of $p_z$, as
\begin{align}
\epsilon_{H,L} = \frac{1}{2}\left(p_0 \pm |p_z|\sqrt{1-\frac{4m_q^2}{p_0^2-p_z^2}}\right),
\end{align}
with $H$ and $L$ corresponds to the upper and lower signs before $|p_z|$, respectively.
From the expression of Eq.~\eqref{eq:DPR:Final}, one immediately notices that in the massless limit $m_q\rightarrow 0$, the rate vanishes. The reason for such a consequence is a mechanism~\cite{Hattori:2020htm} similar to the ``helicity suppression" present in the lepton decay of charged pions.
%----------------------------------------
%.      section
%----------------------------------------
\section{Discussions} \label{sec:discussions}
\begin{figure}[!h]
\centering
\includegraphics[scale=0.36]{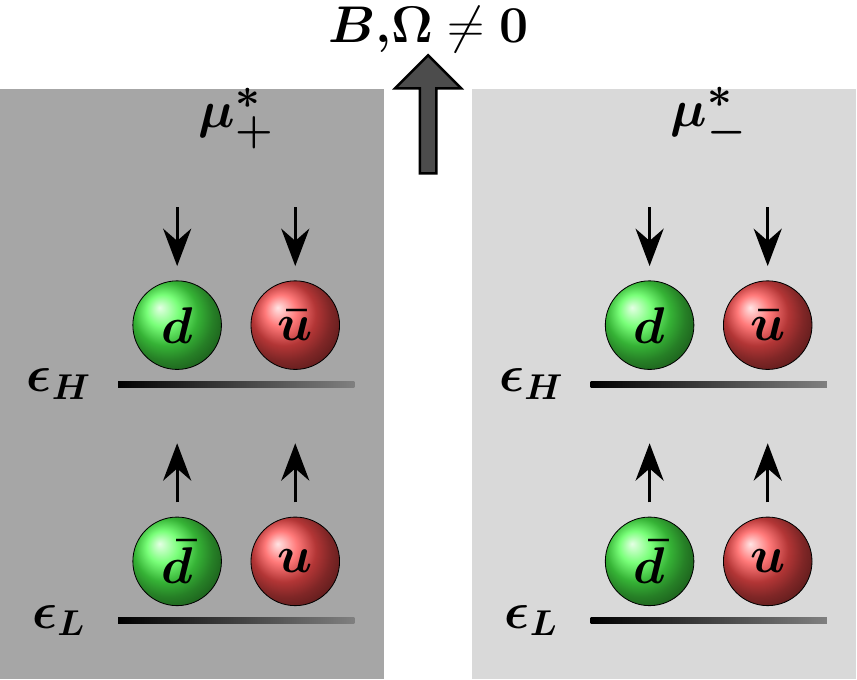}
\caption{[color online] Schematic figures of the two fictitious systems with chemical potential $\mu^{\ast}_{+}$ (left side) and $\mu^{\ast}_{-}$ (right side). The small arrows over the particle symbols represents the $z$-components of spin.}
\label{fig:pic_process_mu}
\end{figure}

To arrive at the simplified result of differential DPR, I have made some assumptions, working best in the early stage of heavy ion collision, under which the validity of our calculation can be justified.\footnote{The typical scale hierarchy that works best in my calculation is $\sqrt{|eB|} > T > \mu > \Omega$.} These are listed below --- 
\begin{enumerate}[(I)]
\item At the initial stage of HIC, the generated magnetic field is very high. It is the highest among all the scales present in the problem. In this scenario, one can assume that the energy gap between the zero-th and first Landau level is so high that the quark antiquark pairs are confined in the LLL~\cite{Hattori:2023egw,Fukushima:2018grm}. Thus, the LLL approximation is justified only in the initial stage.
\item Next, during the early stages of HIC, the radial expansion is not yet developed by particle interactions. In this case one can assume that the quarks are totally dragged by the vortical motion of the fluid. Therefore, the approximation $\bar{\phi}+\Omega\bar{t}=0$ is a good one~\cite{Ayala:2021osy} in this scenario.
\item In my calculation, I have assumed no radial boundary that confines the quarks from escaping the rotating medium. However, to preserve causality, one must have $g_{00} \leqslant 1$. This is ensured by the condition $\Omega R_{\sperp} \leqslant 1$~\cite{Chernodub:2016kxh}. Thus, to secure the validity of our calculation, I have considered the case of slow rotation.
\item When tackling the problem of rotating medium, one should keep in mind that the temperature of the plasma depends on $r_{\sperp}$, the distance of the point in concern from the axis of rotation. Specifically, it changes with distance as $T\left(r_{\sperp}\right)=T\left(r_{\sperp}=0\right)/\sqrt{g^{00}}$, where $g^{00}$ is the zero-th component of metric tensor in curved space time~\cite{Tolman:1930zza}. For the rigid rotation studied here $g^{00}=1-\Omega^2r_{\sperp}^2$. Following~\cite{Braguta:2021jgn}, I have neglected the variation of $T$ with $r_{\sperp}$ by considering it the same with all over the space that at the rotation axis.
\end{enumerate}
After shedding some light on the assumptions made in our calculation, I wish to comment on the translational symmetry in various backgrounds.
\begin{enumerate}
\item $\bm{eB \neq 0}$ \textbf{and} $\bm{\Omega = 0 :}$\\
In the case of non-zero background magnetic field in a non-rotating scenario, the rotational symmetry is explicitly broken in the direction perpendicular to the reaction plane. Due to this explicit breaking, the translational symmetry in one of the coordinate among four space-time coordinates is lost irrespective of the choice vector potential $\mathcal{A}^{\mu}$. The exact coordinate, in which the symmetry is broken, depends on the choice of the background gauge $\mathcal{A}^{\mu}$ used to represent the B-field as shown in the table~\ref{tab:gauge}. 
\begin{table}[!h]
\begin{tabular}{|c|c|c|}
\hline
Gauge choice ($\mathcal{A}^{\mu}$) & Broken Coordinates & Unbroken Coordinates \\
\hline\hline
$B(0,-y,0,0)$ & $y$-coordinate & $t$, $x$, $z$ \\
\hline
$B(0,0,x,0)$ & $x$-coordinate & $t$, $y$, $z$ \\
\hline
$\dfrac{B}{2}(0, -y, x, 0)$ & $r_{\sperp} (=\sqrt{x^2+y^2})$-coordinate & $t$, $\phi$, $z$ \\
\hline
\end{tabular}
\caption{The interdependency between the gauge choice and coordinate in which the translational symmetry is broken.}
\label{tab:gauge}
\end{table}
This loss is reflected by the appearance of a phase factor in the mathematical expression of the propagator
\begin{align}
S_{f, B}\left(\mathcal{X}, \mathcal{X}^{\prime}\right)=\exp \left[i \Phi\left(\mathcal{X}, \mathcal{X}^{\prime}\right)\right] S_{f, B}\left(\mathcal{X}-\mathcal{X}^{\prime}\right),
\end{align}
where $\Phi$ is called the Schwinger Phase factor. The total phase factor becomes $1$ for the diagrams where the fermions appears only in the form of a closed loop~\cite{Chyi:1999fc} (for example in 1-PI diagrams such as photon polarization tensor diagram, triangle diagrams etc.). 
\item $\bm{eB = 0}$ \textbf{and} $\bm{\Omega \neq 0 :}$\\
In the absence of background $B$ field in a rotating medium, the angular velocity $\Omega$, generated perpendicular to the reaction plane, breaks rotational symmetry. However, the system possesses invariance under rotations in the $x-y$ plane. Therefore, in this case, the most natural choice of the coordinate system to work with is the cylindrical polar coordinate due to the geometry of the system. The translational symmetry remains intact along $t, \phi$ and $z$ direction but it is broken in radial directions. This is very evident from the following argument---
consider two observers sitting on a rotating plate at the same radial coordinate but with different angles will agree on some measurement. But in the opposite scenario, i.e. with same azimuthal but different radial coordinate, they will differ. From the intuition of classical mechanics, it is clear that they will have different radial velocity.

\item $\bm{eB \neq 0}$ \textbf{and} $\bm{\Omega \neq 0 :}$\\
For a rotating plasma in the presence of B-field, one can expect the non-breaking of translational symmetry only in two coordinates. However, in our case the translational symmetry is lost only in one direction, thanks to the fact that both of the $B$-field and $\Omega$ are pointing along the same direction.
Now, few things can be noted from the factor in the propagator that breaks translational invariance.
\begin{itemize}
\item In the presence of background B-field and rotations, the factor that breaks translational invariance, denoted by $\aleph$, is of the following form 
\begin{align}
\aleph\left(\bar{t},\xi,\bar{\phi};\xi^{\prime}\right) = \exp\left[-\Xi\left(\bar{t},\xi,\bar{\phi};\xi^{\prime}\right)\right],
\end{align}
where
\begin{align}
\Xi\left(\bar{t},\xi,\bar{\phi};\xi^{\prime}\right) = \frac{\xi+\xi^{\prime}}{2}-\sqrt{\xi\xi^{\prime}}e^{i(\bar{\phi}+\Omega \bar{t})}\quad\text{with}\quad\bar{t} = t -t^{\prime},\,\bar{\phi} = \phi -\phi^{\prime}.
\end{align}
\item Due to the factor of $S(\mathcal{X},\mathcal{X}^{\prime})S(\mathcal{X}^{\prime},\mathcal{X})$ in one loop calculations, the translationaly non-invariant factor, after few steps of algebra, becomes
\begin{align}
&\exp\left[-\Xi\left(\bar{t},\xi,\bar{\phi};\xi^{\prime}\right)-\Xi\left(-\bar{t},\xi^{\prime},-\bar{\phi};\xi\right)\right] = \exp\left[-\left\{\xi+\xi^{\prime}-2\sqrt{\xi\xi^{\prime}}\cos\left(\bar{\phi}+\Omega\bar{t}\right)\right\}\right] \nonumber\\
& = \exp\left[-\frac{|q_fB|}{2}\left\{r_{\sperp}^2+{r_{\sperp}^{\prime}}^2-2r_{\sperp} r_{\sperp}^{\prime}\cos\left(\bar{\phi}+\Omega\bar{t}\right)\right\}\right]. 
\end{align}
\item For $\Omega = 0$, expanding the Cosine and using $x=r_{\sperp}\cos\phi, y=r_{\sperp}\sin\phi, x^{\prime} = r_{\sperp}^{\prime}\cos\phi^{\prime}, y^{\prime}=r_{\sperp}^{\prime}\sin\phi^{\prime}$, one gets back the translational invariance in the radial coordinates. Mathematically, it turns out that
\begin{align}
\lim_{\Omega\rightarrow 0}\exp\left[-\frac{|q_fB|}{2}\left\{r^2_{\sperp}+{r^{\prime}_{\sperp}}^2-2r_{\sperp} r_{\sperp}^{\prime}\cos\left(\bar{\phi}+\Omega\bar{t}\right)\right\}\right] = \exp\left[-\frac{|q_fB|}{2}\left\{(x-x^{\prime})^2+(y-y^{\prime})^2\right\}\right].
\end{align}
This result is in accordance with the non-rotating case where the phase factor disappears in the diagrams with closed fermion loops.
\end{itemize}
\end{enumerate}
The expression of DPR can be interpreted as follows:
One can rearrange the Fermi-Dirac distribution part of the result in Eq.~\eqref{eq:DPR:Final} as
\begin{align}
&2-\widetilde{n}\left(\epsilon_{H}-\mu_{s_q}^{\ast}\right) -\widetilde{n}\left(\epsilon_{L}-\mu_{s_q}^{\ast}\right)-\widetilde{n}\left(\epsilon_{H}+\mu_{s_q}^{\ast}\right)-\widetilde{n}\left(\epsilon_{L}+\mu_{s_q}^{\ast}\right)\nonumber \\
& = \left[1-\widetilde{n}\left(\epsilon_{H}-\mu_{s_q}^{\ast}\right)\right]\left[1-\widetilde{n}\left(\epsilon_{L}+\mu_{s_q}^{\ast}\right)\right]-\widetilde{n}\left(\epsilon_{H}-\mu_{s_q}^{\ast}\right)\widetilde{n}\left(\epsilon_{L}+\mu_{s_q}^{\ast}\right)\nonumber\\
&\hspace{4cm}+\left[1-\widetilde{n}\left(\epsilon_{L}-\mu_{s_q}^{\ast}\right)\right]\left[1-\widetilde{n}\left(\epsilon_{H}+\mu_{s_q}^{\ast}\right)\right]-\widetilde{n}\left(\epsilon_{L}-\mu_{s_q}^{\ast}\right)\widetilde{n}\left(\epsilon_{H}+\mu_{s_q}^{\ast}\right).
\end{align} 
The total DPR in the original system can be interpreted as equal to the sum of that in the two fictitious systems -- one with chemical potential of $\mu^{\ast}_{+} = \mu +\Omega/2$ and another with that of $\mu^{\ast}_{-} = \mu -\Omega/2$ (see Fig.~\ref{fig:pic_process_mu}). The quark flavors can be identified by noting that particles with $s_z=\frac{1}{2}$ possesses lower energy than that of spin $s_z=-\frac{1}{2}$ particles. In both of the systems, a $u$($\bar{d}$) with energy $\epsilon_{L}$ and a $d$($\bar{u}$) with energy $\epsilon_{H}$ annihilates with each other to produce a virtual photon. Note that the statistical weight factors $1-\widetilde{n}(E-\mu)$ and $\widetilde{n}(E-\mu)$ corresponds to a particle in the final state and initial state~\cite{Weldon:1983jn}, respectively.\footnote{To get the weight factors involving antipaticles, the chemical potential $\mu$ is replaced by $-\mu$ in the Fermi-Dirac distribution function. These weight factors are called Pauli suppression factor.}
%--------------------------------------
%          section
%--------------------------------------
\section{Results} \label{sec:results}
\begin{figure}[!h]
\centering
\includegraphics[scale=0.345]{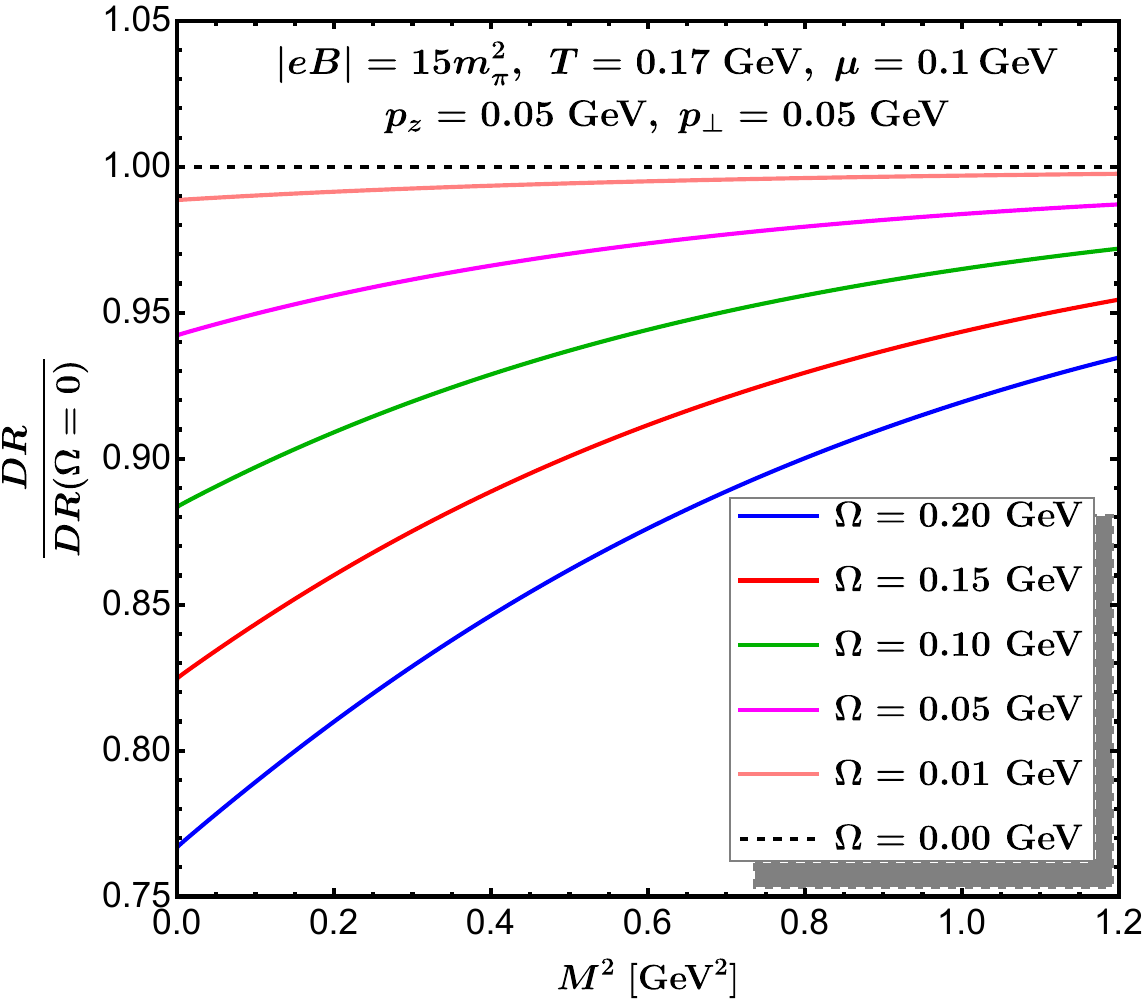}
\includegraphics[scale=0.36]{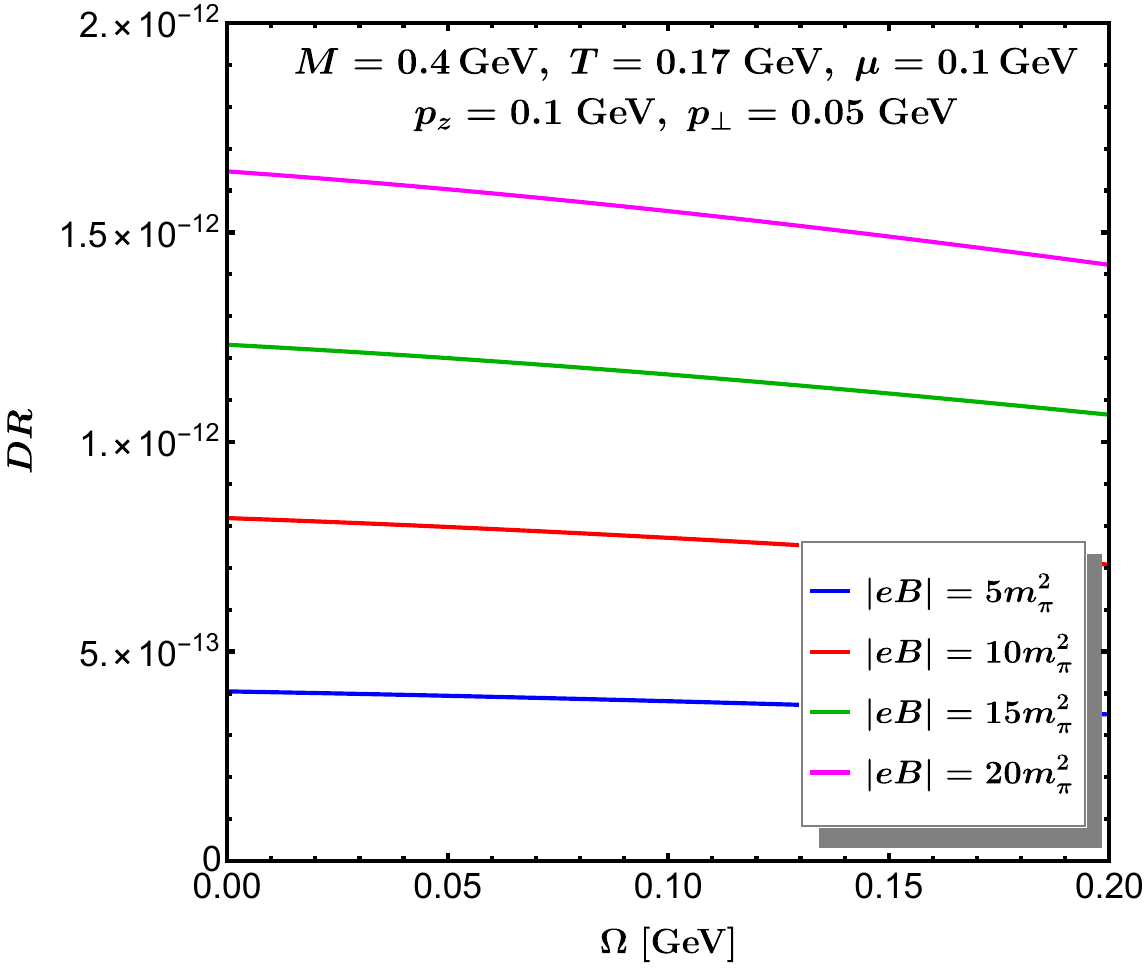}
\caption{[Left Panel] The ratio of DPR at non-zero $\Omega$ and at $\Omega=0$ (along the $y$-axis) for various values of $\Omega$  as a function of the invariant mass of di-leptons $M^2$ (along the $x$-axis). [Right Panel] The DPR as a function of $\Omega$ for $|eB|$ values starting from $5m_{\pi}^2$ to $20m_{\pi}^2$ in steps of $5m_{\pi}^2$.}
\label{fig:DReBOmega}
\end{figure}

For the illustration purpose, I have plotted the dilepton production rate, denoted by DR, with invariant mass squared\footnote{From the conservation of energy momentum valid outside the medium, the di-lepton's invariant mass squared $M^2=(Q_1+Q_2)^2$ is the same as the four momentum square of the virtual photon, i.e., $P^2=M^2$.} $M^2$ at non-vanishing $\Omega$ scaled with DR at $\Omega=0$ for various $\Omega = 0$, $0.01$, $0.05$, $0.1$ and $0.15$ and $0.2$ GeV in the left panel of Fig.~\ref{fig:DReBOmega}. The magnetic field is fixed at $|eB| = 15m_{\pi}^2$, which is reasonably strong in the context of non-central HIC. From the nature of the plot, it is clear that the effect of rotation is to suppress the rate compared to that of non-rotating case. The suppression gets stronger as the invariant mass $M^2$ is lowered. With increasing invariant mass the DPR eventually merge up with the non-rotating strong field case. As $\Omega$ is increased, the DPR increases more rapidly with invariant mass which is clearly seen from the left panel of Fig.~\ref{fig:DReBOmega}.
Next, the behaviour of DPR with $\Omega$ is illustrated in the right panel of Fig.~\ref{fig:DReBOmega} for the values of the magnetic field $|eB|$ starting from $5m_{\pi}^2$ to $20m_{\pi}^2$ in steps of $5m_{\pi}^2$. The rate decreases almost steadily with $\Omega$. However, as the magnetic field is increased, the curve falls more sharply with angular velocity. For example the change of DR with angular velocity is more at $|eB|=20m_{\pi}^2$ than that at $|eB|=5m_{\pi}^2$ in the range $\Omega\in [0,0.2]\;\text{GeV}$. It is evident that the rate increase with magnetic field. The dependence on $|eB|$ on DPR at LLL approximation comes from the term $|eB|\exp (-p_{\sperp}^2/2|eB|)$. This behaviour indicates that with increasing magnetic field the influence of rotation on di-lepton production rate is reduced.\\ 
The result of LLL-approximated case can be recovered by taking $\Omega\rightarrow 0$ limit for $\Omega\neq 0$. However, the $|eB|\rightarrow 0$ limit can not be taken since this calculation is valid only for high $|eB|$. 
\begin{figure}[!h]
\centering
\includegraphics[scale=0.36]{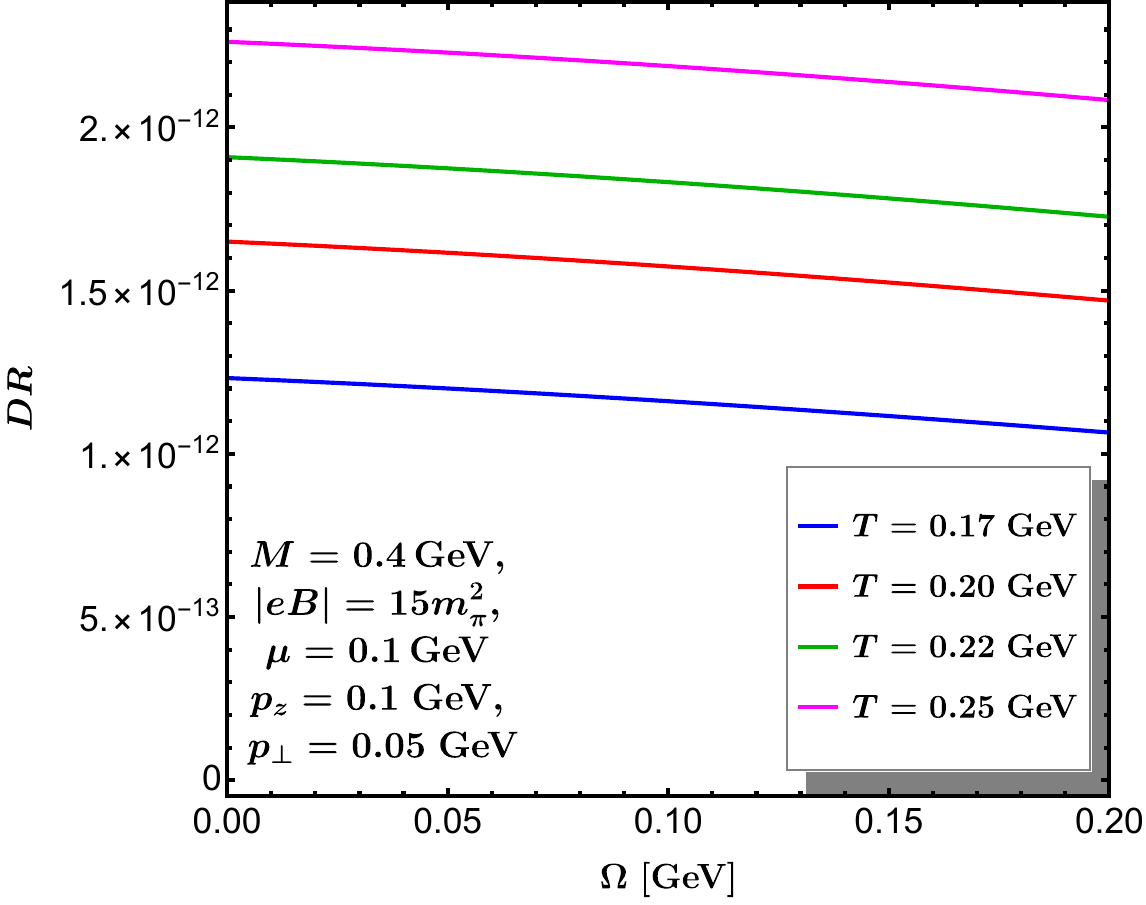}
\caption{The $x$-axis and $y$-axis shows the angular velocity $\Omega$ and DPR, respectively.}
\label{fig:DRTOmega}
\end{figure} 
Finally, the plot of dilepton production rate as a function of $\Omega$ for different temperature of the plasma starting from $T=0.17$ GeV to $0.25$ GeV is displayed in Fig.~\ref{fig:DRTOmega}. Expectedly, the rate increases with temperature. At this point, one can observe that the rate decreases monotonically with $\Omega$ for various temperatures in a similar fashion.\\

The choice of the parameter value $\Omega$ deserve some comments. The vorticity created in the medium depends on the C.M energy ($\sqrt{S_{\textsf{NN}}}$) and the impact parameter (usually denoted $b$) of the two colliding nuclei. The STAR collaboration measured the vorticity $\sim 6$ MeV from $\sqrt{S_{\textsf{NN}}}$ averaged polarization. Furthermore, the AMPT~\cite{Jiang:2016woz} and HIJING~\cite{Deng:2016gyh} predicted it $\sim 27.6$ MeV and $\sim 20$ MeV, respectively. However, in Fig.~\ref{fig:DReBOmega} and \ref{fig:DRTOmega}, I have kept $\Omega$ flexible so that its influence on the rate can be compared. This was done purely from theoretical perspective.

%--------------------------------------
%          section
%--------------------------------------
\section{Conclusion} \label{sec:conclusion}
In this work, I have calculated an analytical expression of dilepton production rate from a rotating color deconfined QGP medium in the presence of strong background magnetic field. I have considered the case of $\bm{B}\parallel\bm{\Omega}$, which is suitable in the context of heavy ion collision experiments. To achieve this goal, I have reformulated the expression of quark propagator in expressed in terms of four-position basis to four momentum basis under the approximation that fermions are totally dragged by the vortical motion of the created medium. In this reformulation of propagator, I only worked with lowest Landau level approximation which is valid when the generated magnetic field is stronger than all the scales present in the theory. In LLL, only spin up $u$, $\bar{d}$ quark and spin down $\bar{u}$, $d$ can propagate. The effect of rotation can be interpreted as an alteration of the quark chemical potential of the system which depends on the charge of the quark under consideration. The dilepton production rate is suppressed compared to the non-rotating due to the presence of rotation. When the invariant mass is reduced, the rate of suppression is found out to be more rapid for higher values of $\Omega$. With increasing $|eB|$, the rate gets enhanced in this calculations.\\

As an outlook one can take a more pragmatic approach to study the DPR by lifting the first three approximations mentioned in Sec.~\ref{sec:discussions}. One such approach is to relax the condition that both the quark and antiquark is in LLL. In this circumstance, it is reasonable to expect that there will be contributions to the rate due to the transition of quarks (anti-quarks) between various Landau levels like $\Omega = 0$ case~\cite{Das:2021fma,Wang:2022jxx}. Also, one can take magneto-rotating system bounded in a cylindrical box whose radius is of the order of the size of the system ($R\sim 10$ fm). Then, one must take into a boundary condition (e.g, MIT boundary condition~\cite{Chernodub:2016kxh}, global boundary condition~\cite{Sadooghi:2021upd}) to strictly ensure that causality in the calculation. Lastly, one can reformulate the rate-formula of DPR from the very beginning by employing the tools of basic thermal quantum field theory when the initial quarks are not in momentum eigenstates. In this direction, one can utilize the skill of QFT in non-trivial backgrounds and consult the classic paper of Weldon~\cite{Weldon:1990iw}. 
%----------------------------------------
%.      Acknoledgement
%----------------------------------------
\begin{acknowledgements}
I would like to acknowledge School of Physical Science, NISER, Bhuwaneshwar.
\end{acknowledgements}
%----------------------------------------
%.      Appendix
%----------------------------------------
\appendix
%----------------------------------------
%.      section
%----------------------------------------
\section{Fermionic Frequency Sum}\label{app:freq_sum}
The frequency sum of interest is written as
\begin{align}
\mathcal{F} \equiv T\!\sum_{k_0}\frac{1}{\left[(k_0 + a)^2-E_1^2\right]}\frac{1}{(p_0-k_0-\bar{a})^2-E_2^2}. \label{eq:app:def_mathcalF}
\end{align}
In the ITF formalism of thermal field theory, the fermionic momentum at finite density is replaced by $k_0 \rightarrow i(2 l+1)\pi T + \mu = \widetilde{\omega}_l + \mu$ and $p_0 \rightarrow i2\pi m T = \omega_m$. Thus, the frequency sum can be rewritten as
\begin{align}
\mathcal{F} &= T\!\sum_{l = - \infty}^{\infty} \frac{1}{\left[(i\widetilde{\omega}_l +\mu + a)^2-E_1^2\right]}\frac{1}{(i\omega_m-i\widetilde{\omega}_l -\mu-\bar{a})^2-E_2^2}\nn
&=\sum_{r_1=\pm 1}\sum_{r_2= \pm 1} \frac{r_1r_2}{4E_1E_2}T\!\sum_{l = - \infty}^{\infty}\frac{1}{i\widetilde{\omega}_l + \mu + a -r_1E_1}\frac{1}{i(\omega_m-\widetilde{\omega}_l) -\mu - \bar{a} -r_2E_2}. \label{eq:app:def_mathcalF_r1r2}
\end{align}
The fraction in the right hand side of the above equation can be written as
\begin{align}
\frac{1}{i\widetilde{\omega}_l + \mu + a -r_1E_1} &= \tilde{n}^{\splus}\left(r_1E_1-a\right)\int\limits_{0}^{\beta} d\tau_1\,e^{-\tau_1\left(i\widetilde{\omega}_l + \mu + a -r_1E_1\right)}\nn
\frac{1}{i(\omega_m-\widetilde{\omega}_l)-\mu - \bar{a} -r_2E_2} &= \tilde{n}^{\sminus}\left(r_2E_2+\bar{a}\right)\int\limits_{0}^{\beta} d\tau_2\,e^{-\tau_2\left[i(\omega_m - \widetilde{\omega}_l) - \mu - \bar{a} -r_2E_2\right]}.\label{eq:app:trick_fsum}
\end{align}
where $\beta \equiv T^{-1}$, $\tilde{n}^{\scriptscriptstyle{\pm}}(E)\equiv \left[e^{\beta(E\mp\mu)}+1\right]^{-1}$.
Thus, substituting Eq.~\eqref{eq:app:trick_fsum} in Eq.~\eqref{eq:app:def_mathcalF_r1r2}, the sum is written as
\begin{align}
\mathcal{F} &= \sum_{r_1,r_2=\pm 1} \frac{r_1r_2}{4E_1E_2}\,\tilde{n}^{\splus}\left(r_1E_1-a\right) \tilde{n}^{\sminus}\left(r_2E_2+\bar{a}\right)\int\limits_{0}^{\beta}d\tau_1\,d\tau_2 e^{-i\tau_2\omega_m}e^{-\mu(\tau_1-\tau_2)}e^{-\tau_1 a + \tau_2 \bar{a}}\nn
&\hspace{8 cm}\times e^{\tau_1 r_1 E_1 + \tau_2 r_2 E_2}\,T\!\!\!\sum_{l = - \infty}^{\infty}e^{-i\widetilde{\omega}_l (\tau_1-\tau_2)},
\end{align}
Next, I employ the identity 
\begin{align}
T\!\!\!\sum_{l = - \infty}^{\infty}e^{-i\widetilde{\omega}_l (\tau_1-\tau_2)} = \delta (\tau_1 - \tau_2)
\end{align}
and perform the integration over delta function to get
\begin{align}
\mathcal{F} = \sum_{r_1,r_2=\pm 1} \frac{r_1r_2}{4E_1E_2}\,\tilde{n}^{\splus}\left(r_1E_1-a\right) \tilde{n}^{\sminus}\left(r_2E_2+\bar{a}\right)\int\limits_{0}^{\beta}d\tau_1 e^{-\tau (i\omega_m + a - \bar{a} -r_1E_1 - r_2E_2)}.  
\end{align}
Performing the $\tau_1$ integral and using $e^{i\beta \omega_m } = 1$, I get after simplifying 
\begin{align}
\mathcal{F} &= \sum_{r_1,r_2 =\pm 1} -\frac{r_1r_2}{4E_1E_2}\,\tilde{n}^{\splus}\left(r_1E_1-a\right) \tilde{n}^{\sminus}\left(r_2E_2+\bar{a}\right)\frac{e^{\beta (r_1E_1-a - \mu)}e^{\beta (r_2E_2 +\bar{a} +\mu)}-1}{i\omega_m -r_1E_1 - r_2E_2 + a - \bar{a}}, \nn
&= \sum_{r_1,r_2 =\pm 1} -\frac{r_1r_2}{4E_1E_2}\frac{1-\tilde{n}^{\splus}\left(r_1E_1-a\right)-\tilde{n}^{\sminus}\left(r_2E_2+\bar{a}\right)}{i\omega_m -r_1E_1 - r_2E_2 + a - \bar{a}}.
\end{align}
At last, continuing analytically to the real value of energy, $i\omega_m \rightarrow p_0 + i\varepsilon$, I get
\begin{align}
\mathcal{F} &= \sum_{r_1,r_2 =\pm 1} -\frac{r_1r_2}{4E_1E_2}\frac{1-\tilde{n}^{\splus}\left(r_1E_1-a\right)-\tilde{n}^{\sminus}\left(r_2E_2+\bar{a}\right)}{p_0 -r_1E_1 - r_2E_2 + a - \bar{a} + i\varepsilon}. \label{eq:app:freq_I_final}
\end{align} 
\section{Kinematics at lowest Landau levels} \label{app:kin}
Here I investigate kinematics of various processes in LLL. I start with the expression of $k_{z}^{(\pm)}$ and quote it again 
\begin{align}
k_{z}^{(\pm)} = \frac{1}{2}\left( p_z \pm \frac{\sqrt{p_0^2\,(p_0^2-p_z^2)(p_0^2-p_z^2-4 m_q^2)}}{p_0^2-p_z^2}\right).
\end{align}
\begin{enumerate}
\item For the production of di-leptons, one must have 
\begin{equation}
p_0 > 0,
\end{equation} 
giving 
\begin{align}
&k_{z}^{(\pm)} = \frac{1}{2}\left( p_z \pm \frac{p_0}{p_0^2-p_z^2}\sqrt{(p_0^2-p_z^2)(p_0^2-p_z^2-4 m_q^2)}\right), \label{eq:kz_sol_2}\\
&k_{z}^{(\pm)}-p_z = -\frac{1}{2}\left( p_z \mp \frac{p_0}{p_0^2-p_z^2}\sqrt{(p_0^2-p_z^2)(p_0^2-p_z^2-4 m_q^2)}\right)=-k_{z}^{(\mp)}, \\
&\epsilon_{k_z^{(\pm)}} = \frac{1}{2}\left(p_0 \pm \frac{p_z}{p_0^2-p_z^2}\sqrt{(p_0^2-p_z^2)(p_0^2-p_z^2-4 m_q^2)}\right), \\
&\epsilon_{k_z^{(\pm)}-p_z} = \frac{1}{2}\left(p_0 \mp \frac{p_z}{p_0^2-p_z^2}\sqrt{(p_0^2-p_z^2)(p_0^2-p_z^2-4 m_q^2)}\right).
\end{align}
\item Now, for the energy value to be real, 
\begin{align}
\left[\left( p_0^2-p_z^2 \geqslant 0 \right) \land \left( p_0^2-p_z^2 \geqslant 4m_q^2 \right)\right] \lor \left[\left( p_0^2-p_z^2 \leqslant 0 \right) \land \left( p_0^2-p_z^2 \leqslant 4m_q^2 \right)\right].
\end{align} 
\begin{enumerate}[(I)]
\item Now, if the condition in the left side of $\textsf{or}$ is satisfied, the $\textsf{and}$ forces one to choose $p_0^2-p_z^2 \geqslant 4m_q^2$. Hence, $p_0 > 0$ implies  
\begin{align}
p_0 \geqslant \sqrt{p_z^2+4m_q^2} \quad\quad \text{Kinematic Region-I}
\end{align}
\item Similarly, if the condition in the right side of $\textsf{or}$ is satisfied, the $\textsf{and}$ forces one to choose $p_0^2-p_z^2 \leqslant 0$. This will imply $(p_0 \geqslant -p_z) \land (p_0 \leqslant p_z)$ for $p_z > 0$ case. For $p_z < 0$, I have $(p_0 \leqslant -p_z) \land (p_0 \geqslant p_z)$. It is because for both sign of $p_z$ the other condition cannot be met simultaneously. Combining both conditions with $p_0 > 0$ gives 
\begin{align}
0 \leqslant p_0 < |p_z| \quad\quad\quad \text{Kinematic Region-II}
\end{align}
\item Lastly, $p_0\geqslant\sqrt{p_z^2+4m_q^2}$ implies 
\begin{align}
&\sqrt{1-\frac{4m_q^2}{p_0^2-p_z^2}} \leqslant 1, \label{eq:B1}\\
&p_0 > |p_z|. \label{eq:B2}
\end{align}
From the conditions in Eq.~\eqref{eq:B1} and ref.~\eqref{eq:B2}, one can write
\begin{align}
p_0 > |p_z| \geqslant |p_z|\sqrt{1-\frac{4m_q^2}{p_0^2-p_z^2}}\;.
\end{align} 
In the region $p_0\geqslant\sqrt{p_z^2+4m_q^2}$, it implies that $\epsilon_{H} > 0$ and $\epsilon_{L} > 0$.
\end{enumerate}
\end{enumerate}
%----------------------------------------
%.      Bibliography
%----------------------------------------

\end{document}